# Recommendations and tools to enable reproducibility in 2D materials research

**[1*]Peter Bøggild, [1]Timothy John Booth, [1]Bjarke Sørensen Jessen, [1]Abhay Shivayogimath, [1]Nolan Lassaline, [2]Stephan Hofmann, [2]Oliver Burton, [3]Kim Daasbjerg, [4]Anders Smith, [5]Kasper Nørgaard, [6]Amaia Zurutuza, [7]Terrance Barkan, [8]Andrew J. Pollard**

[1]DTU Physics, Department of Physics, Technical University of Denmark, 2800 Kgs. Lyngby

[2] Engineering Department, Trumpington Street, Cambridge CB2 1PZ, University of Cambridge, United Kingdom

[3] Novo Nordisk Foundation (NNF) $CO_2$ Research Center, Interdisciplinary Nanoscience Center, Department of Chemistry, Aarhus University, Gustav Wieds Vej 10C, 8000 Aarhus C, Denmark

[4] Villum Foundation, Tobaksvejen 10, 2860 Søborg, Denmark

[5] Novo Nordisk Foundation, Tuborg Havnevej 19, 2900 Hellerup, Denmark

[6] Graphenea S.A., Paseo Mikeletegi 83, 20009 Donostia-San Sebastián, Spain

[7] The Graphene Council, Manchester, United Kingdom

[8]National Physical Laboratory, Hampton Road, Teddington, TW11 0LW, UK

*pbog@dtu.dk

**ABSTRACT**

Research on 2D materials has achieved significant milestones and fuelled a rapidly growing industrial sector. This progress, however, is accompanied by challenges in reproducibility, arising from the atomic thinness, fragility, and environmental sensitivity of these materials. Subtle variations in methods or materials can lead to drastically different outcomes, undermining reliability and slowing down both scientific and technological advances. At the same time, academic publishing and funding systems continue to place greater value on novelty than on efforts to improve reproducibility. This Expert Recommendation outlines concrete, actions researchers can take to improve reproducibility in 2D materials science. We introduce two tools - STEP (Standardised Template for Experimental Procedures) and ReChart (Reproducibility Charter) - designed to support rigorous documentation and transparent sharing of protocols, failure modes, and raw data. To illustrate the application of STEP, we provide three detailed examples covering key processes in 2D materials research: graphene growth by chemical vapour deposition (CVD) on copper foil, wet transfer of CVD graphene, and dry assembly of van der Waals heterostructures. We offer practical recommendations that spans the full research process and show how researchers can engage constructively with stakeholders across academia, funding, publishing, and industry to create a stronger basis for reproducibility, transparency and trust in the field.

## Introduction

Concerns about reproducibility have gained widespread attention across scientific fields, often framed as a 'reproducibility crisis'. A 2012 Nature study attempted to replicate 53 landmark cancer papers and succeeded in only six cases[1], and a recent follow-up study showed similar results[2]. A 2016 poll of over 1500 scientists found that 70% had failed to reproduce another scientist's experiment and 50% their own[3]. Although physics and chemistry fared better than average, an alarming fraction of physicists and chemists remain frustrated by irreproducible studies. More recently, surveys of biomedical researchers have linked poor reproducibility to publication pressure, selection bias, and small sample sizes[4].

While reproducibility challenges exist across many scientific disciplines[5-7], we previously highlighted a distinct 'reproducibility gap' in graphene and other 2D materials[8]. In this rapidly evolving field, scientific and technological progress is often slowed by research results that are difficult to reproduce, despite appearing robust on first inspection.

Since the isolation of atomically thin graphene in 2004, the 2D materials field has rapidly expanded to include a diverse range of materials[9], heterostructures[10], and derivatives, with proposed applications spanning electronics, energy storage, composites, and coatings. Industrial adoption is advancing, and 2D semiconductors now feature prominently in technology roadmaps[11] from leading semiconductor foundries. This rapid growth and high expectation[12] place pressure on the field to deliver robust, reproducible results at both scientific and technological levels.

This situation has produced a surge of method papers, many of which suffer from the same problems of selection bias and sample sizes as discussed in Ref. [4] . Underlying this is a systemic incentive structure that rewards novelty over reproducibility, shaping not only publication practices but also funding and career trajectories[13].

A key reproducibility challenge is the inherent complexity of 2D materials methods, such as synthesis and van der Waals assembly. The atomically thin nature of 2D materials and their exposed bulk make them highly sensitive to ambient conditions, surface contamination, and minor processing variations. Techniques like chemical vapour deposition (CVD) and heterostructure assembly depend on interrelated parameters, many of them undocumented, hidden, or inconsistently reported. Even seemingly trivial factors, such as tape type, substrate roughness, or humidity, can drastically alter the outcomes. This combination of fragility and complexity demands rigorous documentation and structured methods to ensure the field remains scalable and scientifically robust.

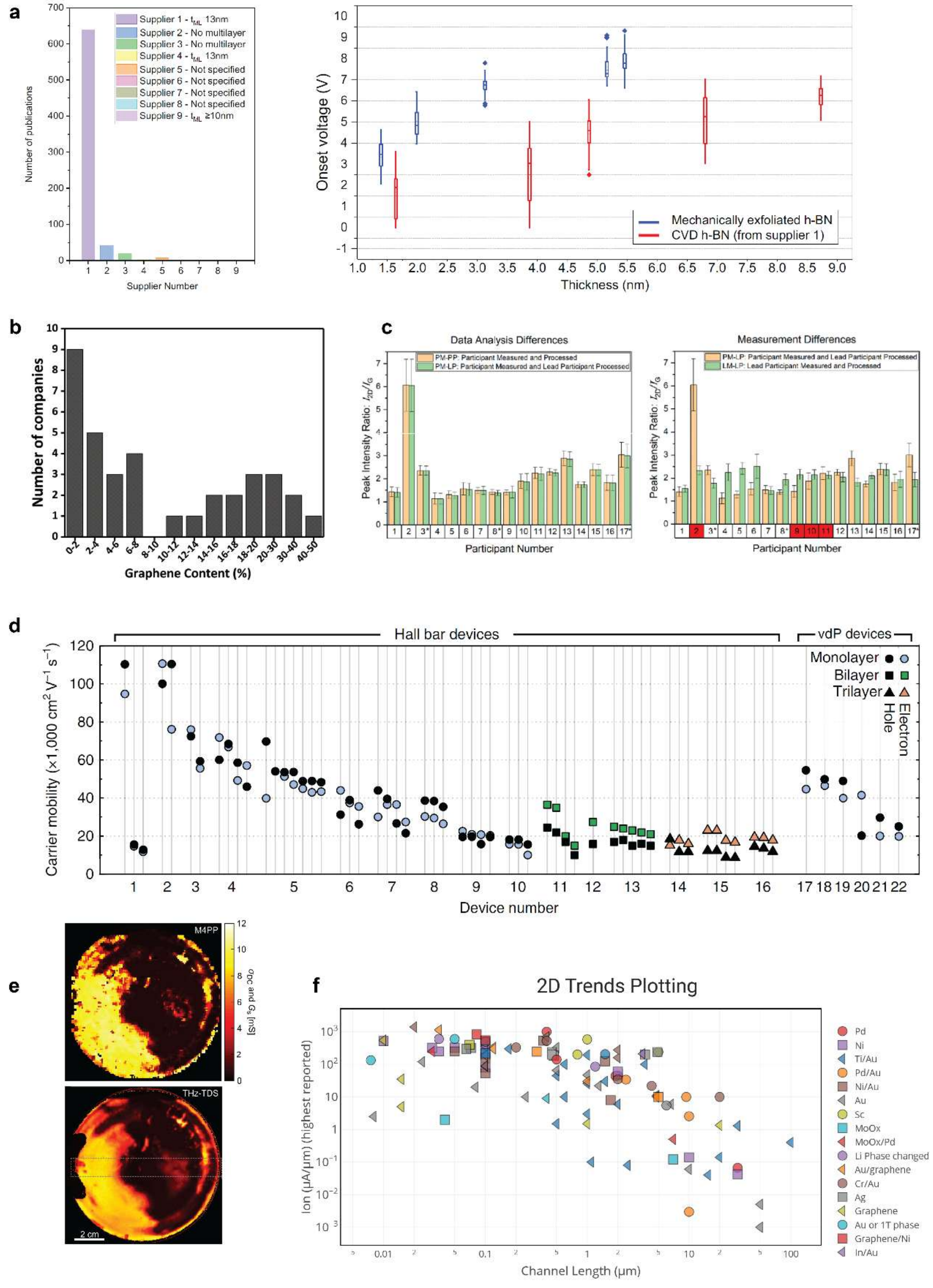

***Figure 1. Reproducibility, repeatability, and validation examples from literature***. *(a) Comparison of CVD samples from commercial hBN vendors and benchmarking against the exfoliated materials. Adapted with permission from Ref [14]. (b) Comparison of the structure and quality of graphene products from 60 commercial liquid-exfoliated graphene samples. Adapted with permission from Ref [15]. (c) Inter-laboratory comparison of the Raman spectroscopy differences of the same graphene samples with respect to the measurement and data analysis. Adapted with permission from Ref [16]. (d) Complete reporting of experimentally measured outcomes of dry stacking method, including negative/substandard results. Adapted with permission from Ref[17], (e) Validation of emerging terahertz time-domain spectroscopy conductivity mapping by benchmarking against industrial micro four-point probe system. Adapted with permission from Ref [18], (f) Online database and visualisation software for benchmarking 2D devices with respect to a range of parameters such as on/off current, channel length, mobility. Adapted with permission from Ref [19].*

Reproducibility has been identified as a critical issue across multiple areas of 2D materials research[20] [15,21-23]. Collaborative inter-laboratory studies have validated key characterisation techniques, including Raman spectroscopy[16] and terahertz time-domain spectroscopy[24]. Beyond characterisation, reproducibility concerns have pointed out for major method families, such as CVD synthesis[25], van der Waals heterostructures[26], electrical contacts, and large-scale transfer[27]. Studies of commercial liquid phase exfoliated graphene [15] and CVD hBN[14] products show significant variability, with many failing to meet their stated specification. Figure 1 illustrates examples that show how systematic comparison, validation, and benchmarking across labs, methods, and materials provide a more reliable foundation for future work than isolated findings.

Large-scale initiatives such as the 2D Experimental Pilot Line (2D-EPL) and 2D-Pilot Line (2D-PL) under the European Graphene Flagship are working to establish robust platforms for wafer-scale 2D graphene device fabrication[28]. These multi-project wafer runs aim to address the reproducible integration of atomically thin materials at scale, one of the most significant challenges in the field.

The present work provides pragmatic approaches to what researchers can contribute, both with respect to their own individual research and in their interactions with the broader research community, represented by key stakeholders such as funding agencies and publishers[29]. These actors play a key role in shaping research culture, where novelty is still often rewarded over robustness. This imbalance must be addressed through coordinated efforts[13] by individuals, institutions, and the systems that evaluate and regulate the funding and dissemination of science.

We offer recommendations for researchers at seven key stages of the research process from planning to follow-up, each aligned with the unique challenges of 2D materials research. To support these efforts, we introduce two complementary tools. The Reproducibility Charter (ReChart) is a structured checklist for declaring reproducibility goals in funding proposals and scientific papers, with a broader scope compared to the solar checklist[30] introduced by Nature Photonics. STEP (Standardised Template for Experimental Procedures) is a detailed reporting format designed to maximise replicability by capturing critical experimental conditions, materials, failure modes, safety and outcomes. The Supplementary Information includes one ReChart example and three full STEP protocols (CVD synthesis of graphene on Cu foil, transfer of CVD graphene from Cu foil, and dry transfer of exfoliated 2D materials) as well as our definitions of reproducibility, replicability, and repeatability. While these two tools have been designed with 2D materials in mind, their use should be straightforward in physical and materials science, and, with some modification, applicable to other natural science and technological disciplines as well.

## Recommendations for researchers

We categorise common reproducibility challenges in 2D materials research across seven stages of the research process: funding, planning, execution, reporting, peer review, citation, and follow-up. For each stage, we offer concrete guidelines and actions that researchers and research teams can take to improve reproducibility of their research, and the chance to convince reviewers, editors, and funding agencies of the robustness and long-term value of a proposed or executed study. Early-career researchers (ECRs) are particularly central to this effort, as they carry out much of the experimental work and shape future research culture. Yet they often face strong pressure to publish quickly, sometimes at the expense of transparency and reproducibility[31]. Senior researchers should lead by example and share clear guidelines on study design, proposal writing, and publication practices. They can also help introduce performance and publication metrics that support ECRs in developing sustainable and reproducible research practices.

*1. Applying for funding*

Although most funding agencies recognise the importance of reproducibility, they often explicitly prioritise novelty and originality over method development and replication. As seen in other fields such as biology and electrical engineering[32], this incentive structure can lead to hyperbole, exaggerated claims and underpowered studies with weak conclusions[33]. Researchers should counter this by explicitly including reproducibility strategies in their proposals—such as allocating budget for method sharing, detailed documentation, and independent validation of results. This might involve letters of support or collaboration with external labs that commit to verifying the methods, either as co-authors on round-robin studies[16,24] or as subcontractors funded by the project. Such proactive measures are likely to be well received by reviewers and funding bodies. While reproducibility is essential for all research, the scope and type of dedicated resources appropriate for a concrete study may depend on the exploratory nature of the project and its technology readiness level (TRL).

*Examples:*

- *(i) Build validation partnerships. Seek letters of support from collaborators who commit to independently test or replicate your methods. Consider including them as co-authors in relevant publications or as subcontractors in the project budget.*
- *(ii) Use reproducibility declarations. Include a ReChart statement in the methodology section of your proposal. Clearly state reproducibility goals, such as sharing of raw data, publishing of negative results, or detailed method documentation.*

*2. Planning a research study*

Poorly defined goals and vague, inadequate methodologies undermine reproducibility from the outset. While a bold, trial-and-error research approach may seem aligned with unexpected discoveries for early career and explorative research, rigorous planning and documentation of experimental procedures are needed to avoid flawed or irreproducible results. Tools like ReChart can support this process by making intended reproducibility effort explicit and part of the plan itself. They help prioritise validation efforts, align team expectations, and identify suitable partners e.g. for round-robin studies[16]. Where standardised or well-established methods exist, they should be clearly identified and adopted. Robust planning also includes data management: outlining how data, metadata, and procedures will be documented, stored, and shared—particularly in collaborative settings. Group leaders should consider that short-term researchers and ECR often feel a stronger need to focus on immediate and individual career goals and may even see transparency as a threat. Timely alignment of expectations can and should prevent later conflicts of interest.

*Examples:*

(i) *Use validated methods. Where possible, adopt standardised or previously validated protocols to reduce uncertainty and improve replicability.*
(ii) *Plan to share procedures. Include in your project plan the intention to publish detailed experimental protocols using STEP, similar to how data management plans are typically included.*
(iii) *Define reproducibility milestones. Set explicit project goals tied to reproducibility, such as "reproducible flake yield in exfoliation demonstrated" or "repeatable domain size in CVD growth validated independently". Link these to specific characterisation methods (e.g. Raman spectroscopy, van der Pauw measurements, optical microscopy).*
(iv) *Align team goals. Make reproducibility, documentation, and transparency part of each team member's role, especially for early-career researchers, and align these responsibilities with their career plan.*

*3. Execution of study*
Reproducibility issues frequently arise from inconsistencies in materials, equipment, and process conditions—especially when parameters are hidden, poorly defined, or inadequately recorded. To address this, researchers should establish systematic baselines and detailed documentation protocols. Tools like ReChart (Table 2) can help structure these efforts, while STEP (Table 1) can support the execution. Comprehensive data logging including both primary parameters (e.g. temperature, gas flows) and secondary factors (e.g. humidity, vibration) [34], can enhance long-term reproducibility, help identify systematic errors, optimise experimental conditions, and reveal underlying mechanisms. Structured digital logs and online protocols that support editing and comments offer a scalable upgrade from personal lab notebooks. Platforms like GitHub and internal wikis are effective for collaborative method development, version control, and transparent revision history. For more structured or data-rich workflows, electronic lab notebooks or spreadsheet-based systems may be better suited to support automation, searchability, and long-term reproducibility. These tools can significantly increase internal consistency and efficiency, while enabling selective sharing within the research team ahead of publication or patenting. At later stages, they also support transparent communication of methods, facilitate peer review, and promote broader knowledge transfer across institutions and the scientific community.

*Examples:*

(i) *Log key parameters automatically wherever possible. Capture both primary (e.g. temperature, gas flows) and secondary (e.g. humidity, vibration) parameters to help identify hidden sources of variation and support long-term reproducibility.*
(ii) *Use STEP to document methods and method development, including parameters, materials, issues, concerns, failure modes, parameter ranges and expected outcomes. STEP protocols can also be used as a structured process development tool to iteratively improve methods, tracking the history and reasons for changes.*
(iii) *Replace personal notebooks with shared digital logs (e.g. spreadsheets, lab notebooks, or cloud-based platforms) to ensure consistent and structured recording of synthesis and characterisation steps. This approach facilitates team-wide access, enables process tracking, and supports expanded data sharing. Include and store annotated images (e.g. flake yield, Raman maps) throughout the process, not just the results.*

*4. Reporting results*
Incomplete or selective reporting of methods and results is a major barrier to reproducibility. Omitting key information, such as polymer type used during transfer, annealing temperature, ambient conditions, or process timing, can prevent others from achieving comparable outcomes.

This is particularly true in sensitive 2D materials research where interface cleanliness critically affects device performance. Even when an experiment is conducted and documented rigorously, the research article may not reflect that rigour for different reasons: space limitations, need to maintain research lead or carelessness. Without this, the scientific community is reduced to the role of an audience second-guessing the tricks of stage magicians, unable to build on published results with confidence. Reporting all results, including those that do not support the hypothesis or even show negative outcomes, will often strengthen the impact and visibility of a published article. Researchers already use supplementary information to provide additional details on their methods or provide extra results, but could further detail the experimental conditions, negative results, challenges, yield, and failure modes. We recommend adopting the STEP (Standardised Template for Experimental Procedures) format, as detailed in the Supplementary Information. In addition, brief tutorial videos and annotated raw data (e.g. Raman maps, AFM scans) can be added as supplementary information or uploaded elsewhere in parallel with the article. Finally, Ph.D. theses remain an invaluable resource for transparent reporting and are well-suited for integrating detailed STEP protocols.

*Examples*

- *(i) Document using a structured template. Use the STEP or another structured format to record detailed methods, including all relevant parameters, known failure modes, process yield, and observed variations.*
- *(ii) Include challenges and negative results. Transparently report unsuccessful trials and known sensitivities, such as flake damage during transfer, polymer residue persistence, or inconsistent nucleation in CVD growth. Note whether these issues were resolved, how, or if they remain open challenges.*
- *(iii) Use multimedia and raw data. Supplement the manuscript with videos, annotated figures, and raw data files, such as Raman maps, AFM scans, or electrical traces, to improve clarity and enable deeper replication.*

*5. Peer review of articles and proposals*

A recurring issue in peer review is the preference for novelty over rigorous methodology—resulting in publications with bold claims but insufficient validation. In 2D materials research, this often manifests as an emphasis on record-breaking properties (e.g. mobility, contact resistance) rather than consistency and reproducibility of experimental techniques such as synthesis, exfoliation, van der Waals assembly or transfer. In addition, publishers' direct interest in maximising the journal impact factor may lead to a certain bias towards letting sensational but unconvincing results pass through peer review, or simply critical reviews to be downplayed in the editorial process. At the same time, reviewers are frequently unable to assess reproducibility due to missing datasets, process parameters, or unclear protocols. Researchers serve a critical role as gatekeepers through participation in peer review, providing the most essential and detailed feedback that many studies and reports receive, with respect to novelty, impact and quality – which includes reproducibility. Reviewers should meticulously and critically assess the methodologies, datasets, experimental and analytical procedures, and weigh this against the scope and type of scientific publication or proposal. Constructive feedback should promote reproducibility by encouraging control experiments, error analysis, and sensitivity checks. Where appropriate, reviewers can also recommend additional statistical or independent validation efforts. Editors, in turn, should moderate these expectations to maintain fairness and avoid excessive burden on authors and reviewers alike.

*Examples:*

*(i)* *Assess reproducibility transparency. Ensure key parameters (e.g. annealing time, humidity, flake alignment), sample sizes, and process variability are reported. For sensitive or novel methods, ask whether independent replication has been attempted or how it is supported (e.g. ReChart, videos, shared data).*

*(ii)* *Recommend detailed protocols. For novel/custom or delicate, complex procedures, such as dry transfer, CVD growth, or twist-angle assembly, encourage authors to provide extended methods in formats like STEP to aid replication.*

*(iii)* *Flag ambiguous reporting. Request clarification if critical details (e.g. measurement methods, exclusion criteria) are unclear. For bold claims of high performance or high-impact, potentially controversial results, suggest stronger controls, error analysis, or replication efforts.*

*6. Citation practices*

While citations and their metrics may not determine research strategies, they may influence them[35]. In addition, citation practices can perpetuate reproducibility issues, for instance, when irreproducible studies continue to be cited more than the studies that question them[36]. This is especially problematic when methodological studies, theses, or protocols published in less prominent journals are overlooked in favour of high-profile but ambiguous reports. Researchers must adopt responsible and critical citation practices to avoid such issues, by evaluating whether a study's methods are transparent, its results independently confirmed, and its reproducibility supported by data. Citing original sources rather than derivative summaries and acknowledging limitations in contested studies strengthens both reproducibility and scholarly integrity. Artificial intelligence (AI) tools, including those implemented by bibliographic databases, journals and third-party companies with research in mind, can support researchers in healthy citation practices, as well as the opposite, that is, automating and even escalating superficial, erroneous, and irrelevant citations.

*Examples:*

*(i)* *Prioritise methodological clarity. Cite studies with well-documented protocols, shared data, and evidence of replication (e.g. STEP documentation, inter-lab validation).*

*(ii)* *Contextualise contested results. Be cautious when citing high-profile studies that are disputed, suspicious, or have underwhelming method descriptions, which could indicate poor replication. If cited, acknowledge the limitations and possible controversies to prevent misrepresentation.*

*(iii)* *Promote best practices. When referencing a study with strong reproducibility practices, e.g. inter-lab validation[16], documented replication, or comprehensive protocols, mention it explicitly to promote good practices.*

*7. Follow-up studies and support*

Follow-up studies are essential for developing more robust practices. For instance, follow-up studies have uncovered that understanding and controlling in-plane and rotational alignment relaxation are crucial for controlling twist angles[37,38]. Too often, however, these are deprioritised due to limited funding, time pressure, or the prevailing emphasis on novelty. After publication of original results, authors can actively support replication of their work by making their data, materials, and/or methodologies openly available to the research community through data repositories or open-access platforms, or prioritising follow-up studies that refine or modify their results. Data management plans have been mandatory in European projects since Horizon 2020, providing clear support for follow-up efforts. Researchers can advocate for and engage in post-publication peer review (PPPR) processes, which continually assess and verify published results, albeit increasing

researchers´ workload [39]. Reward systems such as reproducibility badges or integration into performance metrics could incentivise engagement with PPPR[40,41]. Researchers should also revisit their own work periodically to publish corrections, clarifications, or updates. Inviting peers for short visits or offering online training sessions can also accelerate method transfer. As the authors experienced following the publication of the hot-pickup dry transfer method [8,17], such efforts can dramatically expand the reach and impact of a technique. Similarly, follow-up studies by independent groups can not only validate the original results but also supply practical guidance for replication.

*Examples:*

*(i)* *Open sharing. Share raw data, methods (e.g. STEP protocols), and materials when feasible. Use open-access repositories or institutional archives or attach structured protocols as supplementary files to publications.*

*(ii)* *Support replication. Respond constructively to researchers attempting to reproduce your work. Answering queries, sharing troubleshooting advice, or offering brief virtual training sessions can significantly accelerate replication and community trust.*

*(iii)* *Verify and adapt. Dedicate part of your research to verifying, refining, or adapting previously published methods—including your own—especially when those methods are widely cited or used as a foundation for new work.*

*(iv)* *Correct and update. Proactively review past publications for weaknesses. If errors or uncertainties are identified, update documentation, issue clarifications or addenda, and share lessons learned to avoid perpetuating flawed methods.*

*Table 1. **The STEP framework** breaks down experimental procedures into detailed protocols and recipes, including steps with parameters, materials, issues, and anticipated outcomes. The table overviews the four checkpoints that can be filled out for each process step. Three extensive examples are provided in the SI.*

**STEP - Standardised Template for Experimental Protocols - checkpoint overview**

| Checkpoint | Action | Example | Importance |
|---|---|---|---|
| **PR - Parameters and ranges** | List and record all controlled and uncontrolled parameters. Provide a range of values tested for each critical parameter. | In the growth of $MoS_2$ layers via CVD, controlled parameters might include temperature, pressure, and precursor flow rate. Uncontrolled parameters could involve laboratory relative humidity and ambient temperature. Recommend a plasma power range of 50 to 200 W for PECVD. | Ensures a comprehensive understanding of all factors influencing the experiment, allowing others to replicate conditions accurately. Setting parameter boundaries reduces the risk of failure and increases the likelihood of replicating the results. |
| **ME - Materials and equipment** | Specify all materials and equipment used, including alternatives if primary options are not available. Include details about manufacturers, models, and modifications. | For exfoliating graphene, specify the brand and grade of adhesive tape, the type of mechanical exfoliator, and the properties of graphite flakes. If high-quality CVD graphene is not available, advice is given on where or how to obtain suitable alternatives. | Detailed documentation and alternatives enhance adaptability and prevent variability in replication attempts, ensuring that experiments can be accurately reproduced using equivalent tools and materials. |
| **IWTD - Issues, warnings, troubleshooting and difficulties** | Identify potential safety hazards, operational issues, and provide troubleshooting tips. Point out if and how a step is difficult, and how the experimenter can reduce the difficulty to increase the chance of success. | High-temperature CVD synthesis of $MoTe_2$ with tellurium can lead to the formation of toxic tellurium oxide ($TeO_2$), which is hazardous. For uneven layer thickness in spin coating, adjust the spin speed and solvent viscosity for uniform coatings. | Warning about potential pitfalls and the offer of practical advice improve safety and experimental success rates, helping researchers anticipate and mitigate common challenges, and improving the robustness of methodologies. |
| **VEO - validation & expected outcome** | Describe the expected results or outputs clearly, including any specific observations or measurements that indicate that the process/ characterisation step was successful. | After nitrogen doping of graphene via CVD, expect a noticeable increase in the $I$(D)/$I$(G) ratio in Raman spectroscopy, indicating nitrogen incorporation. XPS should reveal a nitrogen (N 1s) peak around 400 eV, confirming nitrogen's integration into the graphene structure. | The outline of expected outcomes provides a reference for researchers to verify each step of their experiment, ensuring that the results align with theoretical predictions and empirical evidence. |

## Standardised Template for Experimental Procedures (STEP)

Researchers can increase the reproducibility and long-term impact of their results by using standardised protocols and checklists[42,43], providing detailed descriptions of their sample fabrication protocols, experimental procedures, and analyses, and linking these to relevant standards. To facilitate this, we developed the Standardised Template for Experimental Procedures (STEP); a structured format that captures essential parameters, equipment, failure modes, and validation steps. STEP is particularly well-suited to the dynamic and technically demanding nature of 2D materials research, where small variations can have large effects.

By guiding researchers to document their work thoroughly, STEP supports replication, reduces wasted effort, and increases scientific impact. Journals could consider integrating STEP into their submission or review workflows, either as a recommended addendum or as a formal requirement for methods-intensive studies. Similarly, funding agencies could promote or mandate structured reporting tools like STEP to improve transparency and return on investment.

Each STEP protocol breaks the experimental procedure into discrete steps, for which researchers specify: (1) parameters and ranges, (2) materials and equipment, (3) issues, warnings, and troubleshooting, and (4) validation or expected outcomes. Table 1 illustrates these categories with fictious examples inspired by CVD processes, such as graphene and molybdenum disulphide ($MoS_2$) synthesis. Although developed for 2D materials, STEP is broadly applicable across experimental disciplines. Three full examples of STEP protocols are included in Supplementary Information.

A systematic approach to reporting will enhance transparency and trust, encouraging the commercial adoption of new technologies, potentially initiating research on often-overlooked experimental conditions and leading to improved setup design, recipes and scale-up. For researchers engaging in a potentially year-long replication effort to replicate or expand a scientific result, a well-written STEP protocol offers an indispensable starting point. While detailed documentation increases author workload, the time saved for researchers attempting replication could significantly outweigh the additional documentation effort. This is underpinned by the moderate time (3-6 hours) needed to complete each of the extensive STEP examples in the Supplementary Information.

## ReChart: A simple system for declaring reproducibility efforts

We present a reproducibility charter, ReChart, consisting of a list of individual targets to which a study may partially or fully adhere (see Table 2). ReChart can serve as a structured declaration of reproducibility - similar to data availability statements now standard in many journals - and can be included in the main text or in extended form in the Supplementary Information. It also offers a useful reference point in funding proposals and during peer review, enabling more consistent evaluation and communication of reproducibility efforts across research outputs.

*Table 2. The ReChart list of reproducibility targets, with a breakdown of the different types and levels of commitment.*

**ReChart - Reproducibility Charter**

|  |  |
| --- | --- |
| **Replication**: The results have been reproduced by another group using the same methodology (or different methodology) independently verifying the outcomes and demonstrating the reliability of the findings. | a. Repeated by same person<br>b. Repeated by different persons<br>c. Reproduced by another team (same method/materials)<br>d. Replicated by another team (other method/materials) |
| **Detailed methodology:** The experimental methods, techniques, and equipment used in the study are comprehensively described, or previously validated methods are used. This could be achieved with STEP - Standardised Template for Experimental Procedures. | a. Description of methods and equipment used<br>b. Comprehensive description of all steps<br>c. Extended protocol with all parameters, settings, ranges and comments (i.e. STEP)<br>d. Supplementary materials with video demonstrations |
| **Open Data, Materials, and Code Sharing:** Raw data, metadata, and processed data are shared in well-organised and accessible repositories, with clear documentation on how the data were obtained, processed, and analysed. | a. Raw data, device images, or basic code snippets are included<br>b. Data, designs, or code are available on request or hosted<br>c. Well-organized datasets, processing scripts, or fabrication recipes are shared in public repositories<br>d. Data, materials, and code shared in accordance with FAIR* |
| **Error and uncertainty analysis:** A thorough error analysis is included in the research, accounting for uncertainties in measurements, equipment or other aspects of the experimental setup, as well as in data analysis. | a. Basic error analysis included<br>b. Detailed error analysis with uncertainty ranges<br>c. Comprehensive analysis including potential error sources<br>d. Independent verification of error and uncertainty analysis |
| **Negative results reporting:** Negative results and findings that contradict or challenge the initial hypothesis are published with the main result (i.e. in the same article) | a. Acknowledgement of negative results in discussion or SI<br>b. Negative results are reported in detail in article<br>c. Any findings that challenge the hypothesis are analysed<br>d. Publication of a separate study focused on negative results (should be documented, i.e. posted on article repositories) |
| **Robustness and sensitivity assessment:** Information on the robustness and sensitivity of the results is provided, including details on process windows, tolerances, and yield. | a. Basic assessment of method and result robustness<br>b. Detailed process windows, tolerances and yield<br>c. Sensitivity of results to changes in key parameters assessed<br>d. Comparative studies demonstrating the robustness and sensitivity across setups and conditions are included |
| **Open lab policy:** We commit to assisting others in fault-finding attempts to reproduce our results. We maintain an open-door policy, welcoming researchers to visit our laboratory to learn the methodology, observe experimental procedures, and collaborate. | a. Open to email inquiries for fault-finding<br>b. Participation in virtual meetings to discuss methodologies and troubleshooting<br>c. Welcoming researchers for in-lab visits and demonstrations<br>d. Active collaboration to reproduce results (All upon reasonable request) |

*FAIR: Findable, Accessible, Interoperable and Reusable [Scientific Data volume 3, Article number: 160018 (2016) ]

ReChart features seven targets, each with four 'effort levels', as shown in Table 2 and the Supplementary Table 2. While some of these practices may seem self-evident, their systematic declaration helps normalise transparency and sets a shared baseline for reproducibility expectations. By clearly stating which elements a study adheres to, ReChart promotes transparency and quality of both published and planned research studies. An example of a completed ReChart declaration, based on a previously published study[17], is included in the Supplementary Information.

## Recommendations for the broader research community

In the following, we provide brief, targeted recommendations to other stakeholders, most of whom already work actively with or rely on input from researchers, as well as recommendations for what individual researchers can do to engage these stakeholders.

**Private and public funding bodies** should treat reproducibility as a core criterion in grant evaluations and support it with dedicated instruments—such as validation calls, reproducibility supplements, or embedded quality control phases. Funders should also ensure that reproducibility activities are explicitly documented in reports and publications resulting from their grants to align with the goals of providing results with solid fundamental and societal returns on investment. While high-risk, exploratory research must be supported and maintained, funders should always insist on transparent reporting of reproducibility levels and efforts and incentivise replication studies when possible. *Researchers can actively communicate with funders to advocate for reproducibility as a funding criterion by proposing validation plans, requesting dedicated resources and funding calls, and clearly documenting reproducibility efforts in grant applications and reports, directly or through colleagues on boards and panels. Researchers can also engage directly with funders in co-creation of instruments that embrace or prioritise reproducibility efforts and organise workshops with funders. Highlighting own or other researchers' success-stories on open-science initiatives, reproducibility efforts, checklists or other funding agencies who are already responding to the need for better reproducibility.*

**Publishers and Editors** are gatekeepers of scientific communication and have a powerful influence and responsibility for promoting reproducibility within a research field. Currently, the degree to which reproducibility is embedded in editorial guidelines and reviewer assessments varies widely across journals. There are several concrete ways publishers can raise the bar. Journals can request detailed protocols, raw data, and replication disclosures, whether in supplementary materials or via tools like STEP and ReChart. Structured checklists (e.g. Nature's "Reporting Summary", Nature Photonics' solar checklist[30], or ReChart), open peer review formats (as used by eLife), and recognition mechanisms like PLOS ONE's open science badges can all support transparency and help align author and reviewer expectations. Importantly, these measures should be calibrated to the type of journal and article, so as not to discourage timely communication of novel results. At the same time, publishers should consider how editorial practices and reward structures can counteract the pressure to prioritise rapid publication and novelty over transparency and reproducibility. As an example, Nature Human Behaviour introduced 'registered reports', where the journal pre-accepts a study for publication, removing confirmation bias and discrimination of negative results[44]. By recognising replication studies, promoting reproducibility-focused content, and expanding editorial criteria to value methodological rigour, journals can help mitigate the effects of "publish or perish" culture and encourage more balanced academic incentives. *Researchers can engage with publishers and editors by promoting the adoption of reproducibility tools, such as STEP and ReChart, gaining influence by participating in editorial boards or as reviewers, and encouraging the inclusion of reproducibility-oriented content as well as assessment criteria in journals.*

**Research and Technology Organisations (RTOs) and industry** generally play a more indirect role in ensuring reproducibility. Start-ups, SMEs, and larger corporations should be encouraged to partner with researchers early on to validate key findings before committing to scale-up or commercialisation. This can be achieved through formal validation projects with standardisation organisations or metrology institutes, shared benchmarking studies, or participation in pilot lines such as 2D-PL, where methods, materials, and outcomes are openly exchanged. International associations such as the Europe-wide Innovative Advanced Materials Initiative could be instrumental in coordinating and hosting such initiatives. Shared projects that focus on maturing and benchmarking findings should be encouraged, as should industry participation in open-science initiatives, particularly in foundational technology areas. Open Science is a global movement aiming at making research openly available to everyone[45], which can offer a valuable framework for such collaborations, promoting transparency while addressing legal and ethical constraints. *Researchers can initiate reproducibility-focused collaborations with industry partners and RTOs by co-developing benchmarking or validation processes, sharing validated methods, and encouraging joint participation in pilot lines, such as the 2D-PL, or supporting Open Science with industrial or RTO collaborators.*

**Science communicators and journalists** play a pivotal role in shaping public understanding of scientific research and should work closely with researchers to ensure that uncertainty are communicated not as a flaw and weakness, but as a natural and essential part of scientific progress. Media professionals need to recognise importance of independently confirmed studies and, where possible, include reproducibility and validation as a key theme in scientific endeavours. The development of shared codes of conduct for science reporting should be encouraged to bridge the gap between researchers and media professionals. Researchers can build relationships with science communicators and journalists by providing clear, balanced information about uncertainty and reproducibility and promoting accurate storytelling through shared guidelines and open dialogue.

**Trade and professional organisations** are valuable forums for aligning stakeholders around shared standards and practices that support reproducibility. They can connect researchers, industry, regulators, and end-users, and help translate technical results into recognised certifications. For example, The Graphene Council's Verified Graphene Producer programme offers testing services and classification frameworks to support quality and transparency in 2D materials research and industry. Researchers can contribute by engaging directly with these organisations, participating in standardisation activities, certifications, publication of white papers and position statements, while communities such as Versailles Project on Advanced Materials and Standards (VAMAS) can help organising interlaboratory studies and surveys with broader range of partners. By engaging with such organisations, researchers can help shape practical, community-driven norms that reinforce validated science.

**Standardisation organisations (SDOs)** should support the development of consensus-based standards for measurement, analysis, and reporting, helping to translate diverse research practices into broadly accepted and reproducible methodologies in collaboration with the research community. SDOs can further promote reproducibility by aligning standards with widely used instrumentation, offering training programs, and developing digital tools that streamline documentation and implementation. Researchers can participate in standardisation efforts by joining technical committees, contributing empirical insights from their research, develop standards based on suitable research methods[46] and helping align academic practices with measurement, analysis, and documentation standards.

**Curriculum developers** shape the next generation of scientists, yet reproducibility and validation remain largely absent from most academic training relevant to 2D materials research. Educators can address this gap by introducing dedicated courses, faculty seminars, and shared practices around documentation, data sharing, and methodological transparency. Researchers can support these efforts by co-developing teaching materials, integrating reproducibility into thesis requirements, and embedding good practices in lab-based instruction in collaboration with curriculum developers and study leaders. Given their access to teaching programs and mentoring roles, researchers are excellently positioned to help establish reproducibility as a core academic competency.

**Policymakers** are crucial in embedding reproducibility within the broader research ecosystem by integrating transparency and validation criteria into public funding schemes, institutional evaluations, and national research strategies. Initiatives such as the European Commission's Code of Conduct for Research Integrity[47] provide important frameworks, but further efforts are needed to ensure reproducibility becomes a routine expectation across all publicly funded research. Researchers can support this process by contributing expert advice, advocating for evidence-based reproducibility policies, and participating in consultations, advisory boards, and working groups that influence funding guidelines and assessment metrics.

**Challenges and barriers**

In the following we summarise some of the challenges and barriers that could slow down adoption of reproducibility practices and efforts:

**i. Novelty prioritised over reproducibility.** Although reproducibility is essential to scientific progress, novelty continues to dominate publishing practices, funding criteria, and career incentives. Top-tier journals tend to emphasise groundbreaking discoveries rather than replication or validation studies, often assigning lower editorial priority to the latter[5,8,48]. Funding agencies similarly highlight originality and innovation in research evaluations, potentially disadvantaging projects focused on reproducibility. Surveys also show that researchers feel pressured to publish novel findings rather than pursue verification studies[3]. Despite recent efforts to address this imbalance, novelty remains deeply embedded in academic reward structures. As a starting point, journals and funding agencies can implement dedicated sections, grants, and recognition specifically for high-quality replication studies and transparent methodological reporting, and test these via low-risk pilot projects. To overcome this highly complex and deeply rooted issue, however, a coordinated approach will be needed, which calls for partnerships and associations where key stakeholders collaborate on pilot projects, initiatives and standards.

**ii. Inertia towards change among key stakeholders.** Many funding agencies focus on excellence and impact in a rather narrow sense as measured, for instance, by high-profile publications and patents. Publishers may be concerned that introducing detailed reproducibility standards, new metrics and frameworks like STEP and ReChart, could be costly, prolong the review process, reduce journal attractiveness and thus impact factors. There is also a fear of implicitly appearing distrustful towards the research community by highlighting the need for reproducibility. Further, research institutions may be reluctant to implement new reproducibility-focused practices and criteria[49] due to the substantial adjustments needed in established workflows, resources, and training. Such institutional inertia naturally arises when existing systems are well-established, and the immediate benefits or consequences of new methods are not clear. *As mentioned in point (i), the way ahead is through coordinated pilot projects, dialogue and partnerships, involving key research institutions, publishers and funding agencies such as ERC, NSF and leading universities.*

**iii. Too rigid or uniform requirements.** Imposing uniform reproducibility requirements risks slowing innovation, particularly in early-stage low-TRL or exploratory work, where flexibility and creativity are crucial[50]. Also, early-career researchers, who are key drivers of research, already face significant pressure to publish rapidly and demonstrate independence. Increased documentation and reproducibility requirements may disproportionately raise their workload without clear institutional support or incentives, potentially disadvantaging them relative to established researchers. To overcome this barrier, reproducibility efforts should be adaptable to different research contexts, allowing for lighter documentation in early-stage or exploratory projects while encouraging more comprehensive protocols in mature or application-driven work. This flexibility can help maintain creativity and innovation where it is most needed, while still promoting transparency and accountability across the research process.

**iv. Openness as a risk.** Finally, increasing transparency and data openness could conflict with intellectual property protection and commercial interests. Researchers and industry partners may hesitate to share sensitive or commercially valuable information openly, due to concerns about potential loss of competitive advantage or patent opportunities. To address concerns about intellectual property and commercial sensitivity, clear frameworks for responsible data sharing should be established, including options for embargo periods, secure repositories, or selective disclosure. Such measures can help researchers and industry partners share essential information to advance reproducibility, while safeguarding valuable proprietary knowledge and competitive interests.

## Conclusions and Future Directions

The rapid advancement of 2D materials research, technology transfer, and industrial adoption is being slowed by significant challenges related to reproducibility, due to a combination of rapid, non-equilibrium development, strong competition and the intrinsic fragility and sensitivity of atomically thin crystals. This Expert Recommendation does not present a prescriptive roadmap, but rather a pragmatic collection of concrete actions that individual researchers, and the broader research community can adopt to improve reproducibility within 2D materials science and technology.

Our recommendations provide specific guidelines for researchers to enhance the reproducibility of their work and to contribute to the healthy development of the field in this regard. We highlight specific, low-barrier tools that support transparent research practices: a structured protocol format (STEP), and structured reproducibility declaration (ReChart). These tools can be integrated into publications, peer review, and funding proposals, promoting greater methodological clarity and reusability. Importantly, improving reproducibility also requires a reconsideration of how scientific success is incentivised. Current reward structures – the 'game rules' that outline how researchers and Ph.D. students should behave to achieve academic success – often prioritise novelty over reliability and rarely support efforts to validate or confirm results. Shifting these norms is essential for enabling PhD students and experienced researchers alike, to invest in careful, transparent science without compromising career progression.

We also recognise that the need for reproducibility measures is context-dependent, varying by career stage, research field, study type (exploratory or application-driven), and TRL. Sometimes observations and results that are difficult to reproduce can yield important insights and should be reported, and reproducibility requirements should never stand in the way of sharing ideas or preliminary results of interest to the community[50]. However, declaring the degree and nature of reproducibility confirmed or expected nearly always helps avoid misleading readers and supports more robust interpretation of published findings.

We find that given the rapidly growing accessibility and capability of artificial intelligence (AI) in terms of scientific searching and processing; AI-driven tools could significantly boost reproducibility efforts. To make negative result publishing more attractive for journals, AI-agents could systematically compile negative or inconclusive results into accessible databases which could be free or licensed by publishers. Such repositories would enable more effective troubleshooting, predictive experimentation, and better-informed decision-making in laboratories. The effectiveness of AI-training and data mining will rely on better data quality for complex processes such as CVD synthesis, where many parameters are still not well understood. Reproducibility-trained AI-tools could be used to assess manuscripts and proposals pre-publication for researchers, or post-publication. Furthermore, AI-agents could track the adoption and impact of reproducibility frameworks like STEP and ReChart through automated literature analyses, i.e. by introducing reproducibility metrics. This could help researchers, journals, and funding agencies in monitoring the impact of these and other initiatives and adjust accordingly.

The inertia and friction associated with changing established practices are both numerous and complex. For this reason, advancing reproducibility in 2D materials research will ultimately require coordinated efforts across the entire research ecosystem. This consideration led us to propose a stakeholder-specific approach, aimed at building a culture of transparency, methodological clarity, and long-term scientific value within the field. We also anticipate that these recommendations and our analysis are relevant beyond 2D materials, as leading journals and funding agencies rarely focus on this field alone. Meaningful progress will therefore require broader, cross-disciplinary efforts, beyond any single research field.

**Acknowledgements**

P.B. acknowledges financial support from BIOMAG – Novo Nordisk Foundation Challenge Programme and DFF METATUNE. N. L. gratefully acknowledges financial support from the Swiss National Science Foundation (*Postdoc Mobility* P500PT_211105) and the Villum Foundation (*Villum Experiment* 50355). B. S. J. acknowledges financial support from BioNWire – Novo Nordisk Foundation Interdisciplinary Synergy Project. K.D. acknowledges financial support from the Novo Nordisk Foundation $CO_2$ Research Center (grant no. NNF21SA0072700). A. J. P. acknowledges funding from the National Measurement System (NMS) of the Department for Science, Innovation and Technology (DSIT), UK, (Project # 127931).

We acknowledge input and discussions with Alex Wotherspoon, Inge Asselberghs, Rafael Taboryski, Sanna Arpianen and Cedric Huyghebaert.

**Author contributions**

P.B. conceived the study and wrote the first draft of the manuscript. B. S. J., O. B. and A. S. wrote the STEP examples. All authors have read and provided input to the manuscript.

**Competing interests**

The authors declare no conflicts of interest. The opinions expressed in this paper reflect the views of the individual authors and should not be considered as statements of the official policy of their institutions.

# Recommendations and tools to enable reproducibility in 2D materials research

[1*]Peter Bøggild, [1]Timothy John Booth, [1]Bjarke Sørensen Jessen, [1]Abhay Shivayogimath, [1]Nolan Lassaline, [2]Stephan Hofmann, [2]Oliver Burton, [3]Kim Daasbjerg, [4]Anders Smith, [5]Kasper Nørgaard, [6]Amaia Zurutuza, [7]Terrance Barkan, [8]Andrew J. Pollard

[1]DTU Physics, Department of Physics, Technical University of Denmark, 2800 Kgs. Lyngby

[2] Engineering Department, Trumpington Street, Cambridge CB2 1PZ, University of Cambridge, United Kingdom

[3] Novo Nordisk Foundation (NNF) $CO_2$ Research Center, Interdisciplinary Nanoscience Center, Department of Chemistry, Aarhus University, Gustav Wieds Vej 10C, 8000 Aarhus C, Denmark

[4] Villum Foundation, Tobaksvejen 10, 2860 Søborg, Denmark

[5] Novo Nordisk Foundation, Tuborg Havnevej 19, 2900 Hellerup, Denmark

[6] Graphenea S.A., Paseo Mikeletegi 83, 20009 Donostia-San Sebastián, Spain

[7] The Graphene Council, Manchester, United Kingdom

[8]National Physical Laboratory, Hampton Road, Teddington, TW11 0LW, UK

*pbog@dtu.dk

# Supplementary information

## 1. Definition of replicability, reproducibility and repeatability.

Replicability, reproducibility and repeatability are defined in different ways throughout literature, and the definitions used in this work (i.e. in ReChart, main text table 2) follows below.

Scientific reliability relies on the ability to produce consistent results through different forms of experimental verification. Supplementary Figure 1 overviews three key levels of such verification, ranked by increasing demands on effort and resources.

At the foundation lies repeatability, which refers to the ability of the same researcher to obtain consistent results when repeating an experiment under identical conditions. This ensures short-term reliability and internal consistency in a controlled environment.

Moving up, reproducibility involves a different person replicating the results using the same protocol, materials, and equipment. This step tests whether the documentation and methods are robust, transparent, and transferable across different labs, researchers or teams within similar setups.

At the highest level of methodological independence, replicability examines whether a different person can reproduce the findings using different protocols, materials, or equipment. Successful replication under these conditions suggests that the core principles underlying the result are broadly applicable and not dependent on specific experimental conditions.

These distinctions are not merely semantic; they are central to the credibility of scientific knowledge. While repeatability ensures precision, reproducibility safeguards transparency, and replicability validates the universality of the claims.

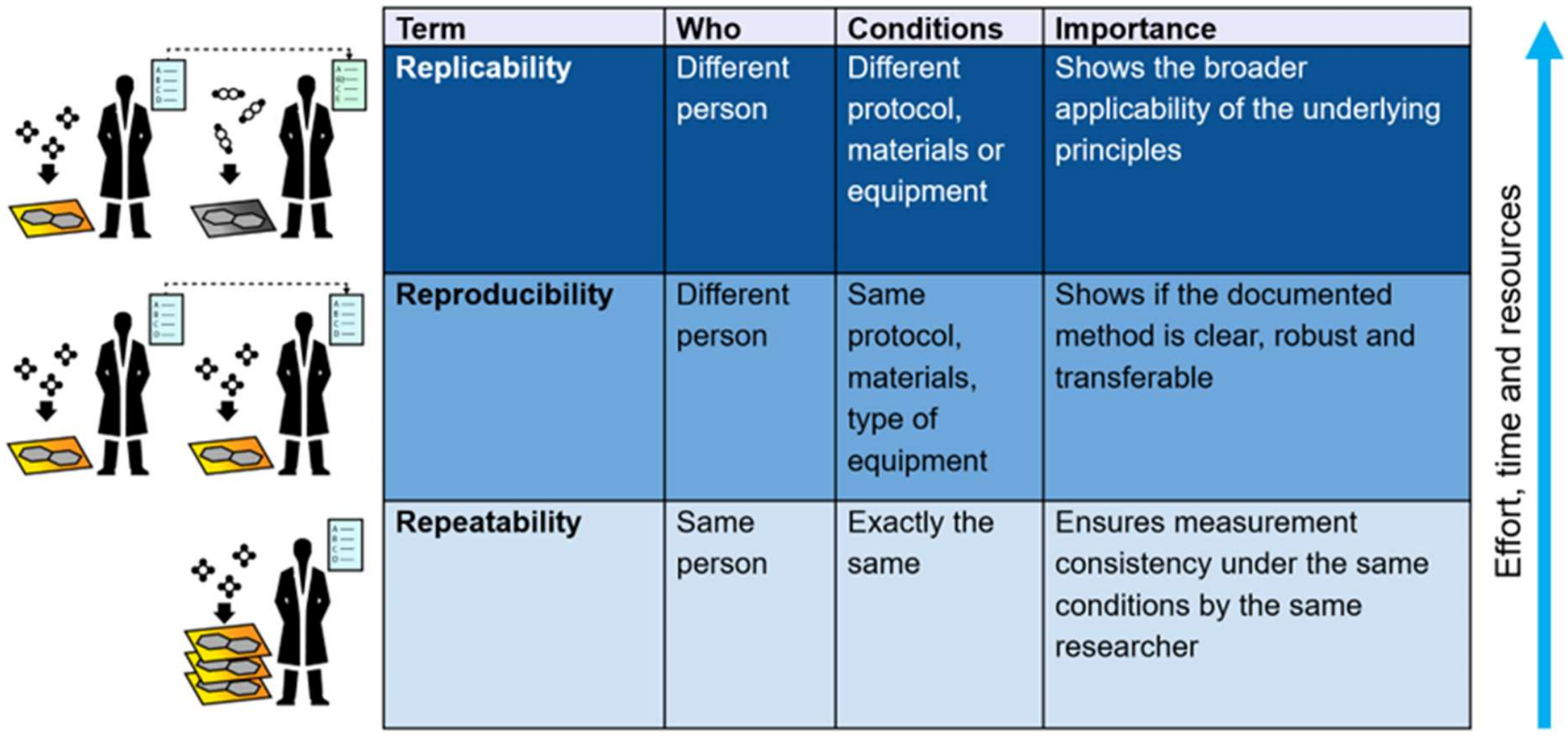

| Term | Who | Conditions | Importance |
| --- | --- | --- | --- |
| Replicability | Different person | Different protocol, materials or equipment | Shows the broader applicability of the underlying principles |
| Reproducibility | Different person | Same protocol, materials, type of equipment | Shows if the documented method is clear, robust and transferable |
| Repeatability | Same person | Exactly the same | Ensures measurement consistency under the same conditions by the same researcher |

*Supplementary Figure 1. Table of distinctions between three essential concepts in scientific methodology used in this work: repeatability, reproducibility, and replicability. Each concept is defined by who performs the experiment, under what conditions, and the significance of each, while the illustrations indicate the principles applied to CVD graphene synthesis. The vertical arrow indicates the increasing effort, time, and resources required as one moves from repeatability to replicability.*

## 2. Example of ReChart declaration for published article

The following shows an example of how the ReChart could be used at the time of publication for a method article[1].

*Supplementary Table 1. Example of ReChart declaration for a published article.*

| | **RECHART declaration of reproducibility effort** | | |
|---|---|---|---|
| | F Pizzocchero, L Gammelgaard, Bjarke S Jessen; Jose Caridad, L. Wang, J. Hone, P. Bøggild, T. Booth, Nature Communications, **7** 11894 (2016) | | |
| **Category** | **Effort** | **Level** | **Description of effort** |
| **Replication** | (a) Repeated by same person | **Achieved** | Two different, complete batches of samples were fabricated, giving similar results. The results are shown in Fig. 6. |
| | (b) Repeated by different people | **Partly** | Three different authors achieved similar stacking results when using the method as described. |
| | (c) Reproduced by different teams (same method/materials) | **None** | |
| | (d) Replicated by different teams (other method/materials) | **None** | |
| **Detailed methodology** | (a) Description of methods and equipment used | **Achieved** | All used methods and equipment are described. |
| | (b) Comprehensive description of all steps. | **Partly** | Details on key steps written in some detail in SI. |
| | (c) Extended protocol with all parameters, settings, ranges and comments (i.e. STEP) | **Partly** | Protocol including some ranges and settings described in SI but not fully, and not in a structured way (such as STEP). |
| | (d) Supplementary materials with video demonstrations | **Partly** | Visual evidence (videos in SI) provided, but not explicitly supplementary video demonstrating every detail of methodology |
| **Open Data, Materials, and Code Sharing** | (a) Raw data, device images or basic code snippets is available on request. | **Achieved** | Standard requirement for Nat. Comm. |
| | (b) Well-organized datasets, processing scripts, or fabrication recipes are shared publicly. | **Partly** | Video recordings showing stacking process in SI. The article and SI contain extensive raw data elements, including Raman spectra, optical and AFM images, electrical traces, and TEM data. Supplementary Movies show the stacking process. Basic declaration of data sharing is included, but not any information on sharing of designs or materials. |
| | (c) Data, materials and code shared in accordance with FAIR. | **None** | |
| | (d) All data, materials, and code are version controlled and | **None** | |
| | include comprehensive metadata for reuse by third parties. | | |
| **Error and uncertainty analysis** | (e) Basic error analysis | Achieved | Statistical errors (mean ± standard deviation) clearly provided for key parameters, e.g., mobility. |
| | (f) Detailed error analysis with uncertainty ranges | Partly | Explicit numerical uncertainties clearly stated for electrical measurements; Raman spectroscopy uncertainties mentioned but less systematically. |
| | (g) Comprehensive analysis including potential error sources | Partly | Clearly identified and discussed potential sources of uncertainty (blisters, temperature effects, doping variability), but no detailed quantitative uncertainty propagation provided. |
| | (h) Independent verification of error and uncertainty analysis | None | |
| **Reporting negative results** | (a) Acknowledgement of negative results in discussion or SI | Achieved | The full range of measured samples, including those with poor performance were shown in Fig 6(a), with statistics. Negative outcomes such as blister formation and flake tearing explicitly acknowledged and briefly discussed. |
| | (b) Negative results are reported in detail in the article | Partly | Negative or problematic findings discussed explicitly (blister formation, failed attempts). The issues with contamination and poor control of drop-down are discussed, leading to variable outcome (as shown in Fig 6). |
| | (c) Any findings that challenge the hypothesis are analysed | Partly | The presence of amorphous carbon in TEM questions the main hypothesis that the drop down technique can squeeze out contaminants, as these do not appear to be squeezed. |
| | (d) Publication of a separate study focused on negative results, linked to or referencing the original article. | None | |
| **Robustness and sensitivity assessment** | (a) Basic assessment of method and result robustness | Achieved | Basic robustness discussed explicitly (variation in process temperature clearly impacts outcome). The number of samples measured is described. |
| | (b) Detailed process windows, tolerances and yield reported | Partly | Detailed windows/tolerances explicitly documented (temperature tolerances carefully explained, blister formation explicitly and quantitatively linked to temperature and processing conditions). The sample yield is partly presented via Fig 6, which only contains finished devices. |
| | (c) Sensitivity of results to changes in key parameters assessed | Partly | Sensitivity explicitly assessed (effects of temperature, contact front speed explicitly demonstrated |
| | (a) Comparative studies demonstrating robustness and sensitivity across setups and conditions included | None | |

| Open lab policy | (a) Open to email inquiries for fault-finding | **None*** (were not declared at time of publication) | *After publication: All email requests (ca 15), were responded to with assistance and guidance, with multiple follow-up conversations to clarify failure modes and give advice. |
| --- | --- | --- | --- |
| | (b) Open to participation in virtual meetings to discuss methods and troubleshooting | **None*** (were not declared at time of publication) | *After publication: We held several online meetings with colleagues to help them reproduce the results) |
| | (c) Welcoming researchers for lab visits and demonstrations | **None*** (were not declared at time of publication) | *After publication: We showed how to use the method to guest researchers and students) |
| | (d) Active collaboration to reproduce results (upon reasonable request) | **None** | |

## 3. Example of STEP for van der Waals heterostructure dry assembly

The example shows the steps needed to produce a van der Waals heterostructure with a monolayer graphene encapsulated in 20-50 nm thick hexagonal boron nitride with minimal bubbles and contamination. In this example we presume that a polycarbonate (PC) stamp has already been made, following for instance the description in Ref [2]. We note that part of the process below (steps 8-9) will depend significantly on the type of polymer and other layers in the stamp stack (there might be several layers on top of the glass slide).

The recipe below is intended for assembly in ambient conditions but could be modified to work in a glovebox. The STEP – Standardised Template for Experimental Procedures is described in the main text.

This example took approximately 4 hours to complete.

*Supplementary Table 2. Example of STEP protocol for van der Waals heterostructure dry assembly.*

| # | Main task | Sub task | ME – Materials and Equipment | PR – Parameters and Ranges | IWTD – Issues, Warnings, Troubleshooting and Difficulties | VEO – Validation and Expected Outcomes |
|---|---|---|---|---|---|---|
| **Description of checkpoints** | | | Specify all materials and equipment used, including alternatives if primary options are not available. Include details about manufacturers, models, and modifications. | List and record all controlled and uncontrolled parameters. Provide a range of values tested for each critical parameter. | Identify potential safety hazards, operational issues, and provide troubleshooting tips. Point out if and how a step is difficult, and how the experimenter can reduce the difficulty to increase the chance of success. | Describe the expected results or outputs clearly, including any specific observations or measurements that indicate that the process/characterisation step was successful. |
| 1 | Exfoliation of graphene | | We used graphite from NGS Naturgraphit. HOPG or most natural graphite providers would server as well.<br>3M Magic scotch tape or blue tape i.e. product 6571 from cleanroomtape.com. The type of tape is not critical, but it may influence the temperature range used later. | | Graphite: there may be differences in mosaic spread, but in our experiences, most will result in sizeable monolayer flakes. | |
| 1.1 | | Fixate tape on work area, | | | | |
| 1.2 | | Press down graphite using gloved fingers or Q-tip, repeatedly. | | | | |
| 1.3 | | Lift graphite gently and press repeatedly on uncovered tape area until high coverage within target area on chip. | | | | High coverage of graphite on tape by visual inspection (before step 4). Microscope not needed. |
| 1.4 | | Cover area with graphite with another layer of tape, to protect it until application to target $SiO_2$ chip. | | | | |
| 2 | Plasma treatment of target $SiO_2$ chip | Position chip in Plasma chamber. Only place 1-2 chips at a time, for quicker handling (see IWTD). We use 30s at low power $O_2$ plasma (20-50 W, at medium-high pressure, 50 mTorr). | Equipment: Plasma ashing system PE-50 from Plasma Etch.<br><br>Material: $SiO_2$/Si: use chip with 90 or 285 nm oxide[3], for high optical contrast. We usually deposit oxide in our own cleanroom on standard Si wafers for exfoliation, using a dry oxidation system. | **Plasma treatment time, power, pressure, chip handling time**<br><br>Parameters may vary considerably depending on system. Optimisation of parameters should be done on local system, to balance coverage and ease of picking up/ stickiness (see IWTD).<br>$SiO_2$/Si: Different thickness may be optimal for other materials[4]. Use high-quality oxide. We recommend using oxide made by dry oxidation. | The purpose is that the treatment suffices to remove hydrocarbons from surface without roughening the $SiO_2$. This balance will be different depending on system and system parameters and would require optimisation.<br><br>It is important to handle the chips relatively quickly, i.e. move to next step within seconds, to avoid unnecessary reabsorption of atmospheric hydrocarbons | |

| 3 | Tape applied to chip | Open the plasma chamber, take chips out on flat working area, open the tape (see 1.4), and press graphite side down onto chip. Rub against top of tape with a blunt object (see ME). | Equipment: blunt object (pencil, a gloved finger) i.e. with a radius of curvature > 3 mm. | **Chip handling time**<br>Time from opening plasma chamber to application of tape should be less than 10 seconds, or as fast as possible, see task 2 (IWTD). | See PR. This step requires a little practice. We recommend be fast but calm. When rubbing tape onto chip, use just enough force to squeeze out trapped air bubbles. | Validation is during optical inspection |
|---|---|---|---|---|---|---|
| 4.1 | Heat treatment | Heat chip on hotplate and remove tape. | Any hotplate with precise temperature control (plus minus 5 degrees) will do. | **Temperature, time**<br>We use 100 C in 1 minute. Time can be increased to 2 minutes (maybe more), but should not be less than 1 min. We recommend precise temperature control. 100 °C is optimal for 3M Scotch Tape to be soft enough to promote conformation of graphene flakes to surface, while not melting. The optimal temperature may be different for other types of tape but should then be kept consistent. | | |
| 4.2 | | Remove tape | 2 pairs of tweezers, one for holding the chip, and one for removing the now soft tape. For high temperatures, use metal tweezers. | We do not note any important differences by varying angle, speed or force at this step. | Don't burn your fingers.<br><br>If the coverage of graphene/graphite is low, considering increasing the plasma time or power. | Optical inspection with microscope: there will be plenty tape residues on the chip, but these should around the deposited graphene/graphite areas, not on top. It may look messy, but it should not matter. |
| 5 | Exfoliation of hBN | The remaining part of the process for hBN is identical to 1-4. | We use either hBN acquired from HQ Graphene[1], or from collaborators at NIMS in Japan[5].<br><br>We use 3M Scotch tape like in step 1. | Tape: some groups recommend using 3M Scotch Greener tape in combination with 3M Scotch tape, but this is in our experience not necessary. | | |
| 6 | Identification and selection of graphene flakes | Typically, mono-, bi- and tri-layers are of interest. | Optical microscope with 100x objective (preferably) and 10x for overview screening of larger areas. | White light source. | It will require practice to consistently discriminate graphene flakes based on layer thickness. This can be solved by using an automated "flake finder" method [ref]. | On a 1 $cm^2$ chip, we expect to find 5-10 monolayer flakes with areas of at least 100 $\mu m^2$. This can vary significantly from chip to chip, and with practice.<br><br>Depending on the application prioritise flakes with (1) no visible damage, cracks or contamination, (2) large straight edges, (3) no folds and wrinkles. |
| 7 | Identification and selection of hBN flakes | Typically, depending on application, flakes ranging from a few (2-4 nm thickness) to many (40-50 nm thickness) layers are of interest. | Same as task 6. | Same as task 6. | (1) it can be hard to see monolayer step edges, i.e. assess the uniformity of hBN due to its low optical contrast.<br>(2) hBN flakes are diffraction-coloured according to their thickness, and we recommend creating a baseline colour map using atomic force microscopy for quick reference.<br>(3) Darkfield microscopy may significantly help to highlight step-edges and structural defects<br>(4) Averaging of several/many images effectively increases the signal-to-noise ratio allowing thinner flakes and smaller defects to be observed. | See IWTD<br><br>On a 1 $cm^2$ chip, we expect to find 5-10 flakes in the 20-50 nm thickness range with uniform areas of at least 100 $\mu m^2$. This can vary significantly from chip to chip, and with practice. |
| 8 | Heterostructure assembly (hBN/G/hBN) | | Key equipment is using a stacking system, consisting most often of an optical microscope equipped with micromanipulators (at least 3 degrees of freedom – XY + Z), clamp or vacuum fixation of sample and chip heater with temperature controller. We use both homebuilt[1] and commercial systems [HQ Graphene '2D Heterostructure Transfer System']. | | It is beyond the scope of this STEP recipe to specify the stacking system; however, a few key concerns include: (1) heater and temperature controller which can reliably control the temperature up to 200 °C with 0.1 °C precision. (2) should be placed in a vibration-free environment or on a vibration isolation stage. (3) Objectives should include (very/ultra) long working | |

| | | | | | distance 20x-50x objective. WD should long enough to accommodate the glass slide between the objective and the silicon oxide chip. Glass-corrected objectives can help increasing clarity of imaging.<br>(4) We recommend using a motorised z-stage, as this greatly simplifies the task of achieving a smooth, continuous approach, as well as achieving reproducible, operator-independent results. | |
|---|---|---|---|---|---|---|
| 8.1 | | Pickup of first hBN flake | | | | |
| 8.1.1 | | Place hBN chip on stage and identify the target flake. Center it in the optical viewfield. | | | | |
| 8.1.2 | | Glass slide mounted on microcontroller with PC area approximately centered around target flake. | | | | |
| 8.1.3 | | Flake chip and slide is brought close to each other at a safe distance | | **Distance**: 1-2 mm. | Smaller distances make accidents more likely. When temperature is increased in next step, expansion can lead to unwanted contact. | |
| 8.1.4 | | Temperature is increased to 110 °C. | | **Temperature, Stabilisation time**<br><br>Wait 1-2 minutes for stabilisation<br><br>Temperature should be 110 °C. Note: this is not important; there can be advantages of using significantly lower temperatures. | This task can be done earlier (i.e. at 8.1.1) | When focusing on the target flake, a change in temperature will lead to drift of the focus. When focus no longer drifts, the system can be taken to be thermally stable. |
| 8.1.5 | | Stamp brought into contact with hBN flake. (1) the polymer stamp will often touch first at a certain point close to the flake, which is clearly observable in the microscope<br>(2) upon further approach, the contact area will expand until its edge has passed over and is now fully covering the hBN flake. | | **Time, rate**<br><br>It is not important to do this very slowly; from initial touch-down to stamp covering hBN (ready to retract) can be a few minutes. | (1) We recommend continuing until the edge of stamp-chip contact area extends at least 50-100 microns beyond the flake. This helps to ensure consistent lift-up in step 8.1.6, even if mechanical or polymer drift occur.<br>(2) While we here focus on the approach done by the micromanipulator, similar results can be achieved by slowly increasing the temperature using the thermal expansion of the polymer to close the gap between stamp and polymer.<br>(3) If the stamp and chip are already in contact, this can give a very smooth approach; care should be taken to use only moderate temperature increase (e.g. 10-20 °C, depending on polymer) | The stamp-chip contact area covers the flake, extending 50-100 microns beyond. |
| 8.1.6 | | Stamp with hBN flake is lifted up / retracted from chip. | | **Stabilisation time, speed of retraction**<br><br>We recommend waiting at least 3-5 minutes for the system to achieve thermal and mechanical stability. In our experience this increases the chances of successful pickup.<br><br>Retraction speed can be similar to or slightly lower than in step 8.1.5. | | Retraction speed can be compared to the approach speed by watching the stamp-chip contact line, or by using a motorised z-stage (see point 8)<br><br>Validation of successful pick-up is done by making sure that the hBN flake is no longer on the chip. It is advisable to inspect the glass slide in a different/good optical microscope (100x) to ensure the structural integrity of the hBN as well as the contact with the stamp. |
| | | | | | | At this stage it is recommendable to record optical images for later use (troubleshooting, optimisation or publication). |
| 8.2 | | Pickup of graphene flake, using the stamp with hBN flake. | As in 8.1 | | | |
| 8.2.1 | | Place graphene chip on stage and identify the target flake. Center it in the optical view-field. | | | | |
| 8.2.2 | | Glass slide mounted on microcontroller with PC area approximately centered around target flake. | | | | |
| 8.2.3 | | Flake chip and slide is brought close to each other at a safe distance | | As 8.1.3 | As 8.1.3 | As 8.1.3 |
| 8.2.4 | | Temperature is increased to 110 °C. | | As 8.1.4 | As 8.1.4 | As 8.1.4 |
| 8.2.5 | | Stamp brought into contact with graphene flake. Align graphene and hBN flakes using microcontrollers. The rest of the step follows 8.1.5. | | **Time, rate**<br>Initially, approach the two surfaces slowly until first contact. From here the approach process should be very slow (typical time from first contact to final state is over 10 minutes). | (1) When aligning the two target flakes, they will initially have different focal planes (until they touch).<br>(2) Continuously compensate the lateral positions of the flakes during vertical approach, as they tend to drift sideways.<br>(3) Gradually move the stamp closer to the chip while performing this procedure. As the targets are very close, the focal plane will be almost identical.<br>(4) The very slow approach allows contaminants and bubbles at the mechanical junction between the flakes to be expulsed. If the flakes are brought into contact too quickly, bubbles and contamination might get trapped. These can still be agglomerated using thermal treatment leaving more space for devices[6], removed by postprocessing[2], scraped by contact-mode AFM[7] or avoided entirely by assembly in vacuum[8]. The advice above will limit the need for either of these when assembling in ambient conditions.<br>(5) If contamination and bubbles are still problematic, increasing the temperature (i.e. up to glass transition temperature of the polymer which for PC is 147 °C) during drop-down can help. | As 8.1.5 |
| 8.2.6 | | Stamp with hBN flake is lifted up / retracted from chip. | | **Stabilisation time, retraction rate**<br>The surfaces are kept in contact for at least 30 minutes, corresponding to the baking step in Ref [1]. This appears to facilitate the adhesion between the graphene and hBN flakes.<br><br>Retraction rate: 1-5 minutes from full contact to full release. | (6) On the retraction rate: Provided the above 'baking' step is successful it should not be necessary to retract very slowly. A time from full contact to release of 1 minute should be sufficient, however, spending more time could be safer and does not pose any other problems we know of. Different group use very different strategies for this step, with seemingly similar outcomes. | |
| 8.3 | | Drop-down on second hBN flake | | | | |
| 8.3.1 | | Place hBN chip on stage and identify the target hBN flake. Center it in the optical view-field. | | | | |
| 8.3.2 | | Glass slide with hBN/graphene stack | | | | |
| | | mounted on microcontroller. | | | | |
| 8.3.3 | | Flake chip and slide is brought close to each other at a safe distance | | As 8.1.3 | As 8.1.3 | As 8.1.3 |
| 8.3.4 | | Temperature is increased to 110 °C | | As 8.1.4 | As 8.1.4 | As 8.1.4 |
| 8.3.5 | | Stamp with stack brought into contact with graphene flake. Align hBN/graphene stack and hBN flake using microcontrollers. The rest of the step follows 8.1.5. | | As 8.2.5 | As 8.2.5 | As 8.2.5 |
| 8.3.6 | | Temperature is increased to 200 C until polymer stamp is melted onto chip, separating it from glass slide. | | **Temperature, Time**<br>Temperature at 200 C ensures that PC reflows (starts at 155 °C).<br><br>The time should be at least 10 minutes, to ensure that the reflowing process has completed. | During the heating process, the polymer will expand, so it is advisable to compensate by gently adjust the z-position (height) to maintain a relatively constant contact area. | PC polymer will not change appearance. |
| 8.3.7 | | Retraction of glass slide | | **Time**<br>Retraction time should be slow enough to allow the separation of the glass and polymer to progress controllably. We recommend 5-10 minutes for this process (from contact to non-contact) as a starting point. | If the stabilisation time is too short, the polymer may be lifted partly up with the glass slide, potentially damaging the stack. | After retraction of glass slide, the stack should be located on the chip, visible through the deposited polymer (8.3.6). |
| 8.3.8 | | Removal of melted polymer is achieved using 1 or more repetitions of the following cycle: (1) Dip chip for 10 seconds in Chloroform, move directly to (2) 10 seconds in acetone, and (3) 10 seconds in IPA.<br><br>Following this, the chip is gently dried with a flow of nitrogen. | 3 beakers, with (1) Isopropanol Alcohol, (2) Chloroform*, (3) Acetone<br><br>Fumehood (see IWTD). | Time per dip. Approximately 10 seconds for each dip. Nitrogen blow until dry. | **WARNING:**<br>(1) Chloroform is toxic and should handle with the greatest care, following all possible safety precautions. The removal process **MUST** be carried out in a fumehood and/or personal protection equipment, and consultancy and approval with the local working safety responsible is mandatory. Stay safe.<br>(2) The reason for the quick alternating dips in three liquids is that we observed that prolonged immersion in chloroform can lead to contamination and degradation of some 2D materials. It is not known whether this is detrimental specifically to hBN/G/hBN stacks, yet we recommend this procedure for caution. | The chip should appear clean upon visual (naked eye) inspection. The cycle (Chloroform, Acetone, IPA) can be repeated several times until the chip is clean. |

## 4. Example of STEP for CVD graphene growth onto Cu foil

This protocol describes a reproducible method for growing polycrystalline monolayer graphene on commercial copper foil using a dual heater commercial Aixtron BM Pro 4" CVD reactor. The procedure represents the integration of standard methods and research on growth from the Hofmann group (e.g. Burton et al. [9], Braeuninger-Weimer et al. [10]) and is designed to maximize repeatability through cleaning, surface pre-treatment, and controlled processing conditions. Here repeatability refers to the nucleation density and growth rates of graphene on Cu foil, as well as 'high quality' as defined by low (<0.03) averaged D/G peak ratios from Raman spectroscopy, the measurement details of which are beyond the scope of this STEP, however we note that graphene transfer is intrinsically linked to the end quality of graphene used. The process also does not describe variations to change, for example, nucleation densities and growth rates, many of which are the subject of numerous systematic studies in the literature.

In brief, the process includes cutting the foil, solvent cleaning (Section 1.2), oxide removal by immersion in glacial acetic acid (Section 1.3), optional electropolishing (Section 1.4) to reduce surface roughness, and a controlled hot-plate oxidation step (Section 1.5) performed on the entire foil. Reactor pre-treatment, calibration, sample loading, annealing (in Ar and $H_2$), graphene growth under controlled gas flows, controlled cooldown, and sample extraction are also described. All steps are performed in a cleanroom environment with appropriate PPE, including clean gloves, hairnets and cleanroom suits. This protocol assumes that in all other respects other than those explicitly mentioned, the CVD reactor is well calibrated (e.g. MFCs, power measurements, heater uniformity), and that the CVD chamber is well maintained (e.g. regular quartz and graphite parts replacement and cleaning, heaters not degraded) as the calibration and maintenance of a CVD system is beyond the scope of this STEP.

This example took 4-5 hours to complete.

*Supplementary Table 3. Example of STEP protocol for CVD graphene growth onto Cu foil.*

| # | Main task | Sub task | ME – Materials and Equipment | PR – Parameters and Ranges | IWTD – Issues, Warnings, Troubleshooting and Difficulties | VEO – Validation and Expected Outcomes |
|---|---|---|---|---|---|---|
| **Description of checkpoints** | | | Specify all materials and equipment used, including alternatives if primary options are not available. Include details about manufacturers, models, and modifications. | List and record all controlled and uncontrolled parameters. Provide a range of values tested for each critical parameter. | Identify potential safety hazards, operational issues, and provide troubleshooting tips. Point out if and how a step is difficult, and how the experimenter can reduce the difficulty to increase the chance of success. | Describe the expected results or outputs clearly, including any specific observations or measurements that indicate that the process/characterisation step was successful. |
| 1 | Catalyst Preparation | | | | | |
| 1.1 | | Section the commercial Cu foil into desired dimensions. Cut the foil into ~5×5 $cm^2$ pieces using clean, sharp shears or scissors. Handle only with tweezers in the corners to avoid contamination and mechanical deformation. | Polycrystalline Cu foil (25 μm thick, e.g. from Alfa Aesar); precision shears; cutting guide/template | Target dimensions: 5×5 $cm^2$; | Handle with care to avoid introducing strain or scratches which could affect crystallographic properties during annealing/growth. Handle with tweezers at the corners, as these regions will be discarded after growth. | Uniform, smooth-edged pieces verified by visual inspection; absence of folds, creases, or pinch marks. |
| 1.2 | | Catalyst Cleaning | High-purity acetone; high-purity IPA; clean glass containers (4" diameter crystallising dishes work well); ultrasonic bath/sonicator; IPA in wash bottle | | The Cu foil must be handled with care throughout to avoid any deformation. | A visually clean, shiny Cu surface with minimal residue confirmed by optical inspection. |

| 1.2.1 | | Prepare clean solvent baths by filling two clean glass containers with ~200 mL each of high-purity acetone and high-purity isopropanol (IPA). Ensure containers are dust-free and rinsed with DI water before use. | | ~200 mL per container; use freshly opened bottles | Ensure containers are free of dust or other contamination to prevent re-contamination. | Solvent baths are clear and free of particulates. |
|---|---|---|---|---|---|---|
| 1.2.2 | | Immerse the Cu foil fully in the acetone bath and sonicate for 2 minutes at room temperature. Monitor the sonicator to ensure temperature remains constant. | Prepared acetone bath; | Sonication time: 2 minutes; temperature: ~25 °C | Avoid prolonged sonication to prevent overheating; ensure foil remains undamaged. | Surface contaminants (oils, grease, dust) begin to be removed; foil appears uniformly darkened by solvent action. |
| 1.2.3 | | Rinse off the acetone by gently spraying the foil with IPA from a wash bottle to minimize residual acetone. | | Rinse duration: approximately 30 seconds | Handle the foil gently to avoid scratching; ensure complete removal of acetone. | Foil shows reduced acetone residue and appears uniformly cleaned. |
| 1.2.4 | | Transfer the Cu foil to the IPA bath and sonicate for 2 minutes at room temperature, ensuring full immersion. Handle the foil carefully to avoid scratches. | Prepared IPA bath; | Sonication time: 2 minutes; room temperature (~25 °C) | Avoid excessive sonication that might damage the surface; ensure complete immersion. | Further removal of residues; foil surface becomes uniformly clean as observed by visual inspection. |
| 1.2.5 | | Immediately dry the Cu foil using a gentle, continuous flow of high-purity $N_2$ gas for 1–2 minutes until visibly dry. | High-purity nitrogen ($N_2$) gas source with clean $N_2$ nozzle | $N_2$ drying: 1–2 minutes; use gentle flow | The foil is placed on some cleanroom wipes for support to avoid deformation. A lower $N_2$ flow is preferable to avoid the risk of catching the foil in turbulence and crumpling it. | The foil is dry and shiny. |
| 1.3 | | Oxide removal (optional) | Glacial acetic acid; DI water; clean glass containers (4" diameter crystallising dishes work well); clean container for rinsing; High-purity $N_2$ gas source | | | Foil is completely dry and exhibits a bright, oxide-free surface as confirmed by visual inspection. |
| 1.3.1 | | Prepare the oxide removal solution by mixing glacial acetic acid with DI water in a 1:1 volume ratio in a clean container. This solution will remove native oxide from both sides of the foil. | | 1:1 volume ratio; solution prepared at room temperature; ~200 mL total volume (adjust as needed) | Acetic acid is corrosive; use appropriate PPE (gloves, goggles, lab coat) and work in a fume hood. | A uniformly mixed acid solution free of particulates. |
| 1.3.2 | | Immerse the entire Cu foil (both sides) in the acetic acid solution for 30–60 seconds to remove the native oxide layer. | Prepared acetic acid solution | Immersion time: 30–60 seconds at room temperature | Ensure complete submersion of both sides; | Visible removal of native oxide; foil appears uniformly metallic (bright, shiny) after subsequent rinsing. |
| 1.3.3 | | Immediately transfer the foil to a DI water bath and rinse thoroughly to remove all acid residues. | | Rinse duration: 2 minutes; gentle agitation recommended | Incomplete rinsing may leave acid residues that can affect subsequent steps; ensure full removal of acid. | Foil is free of acid residues, as confirmed by a uniform appearance. |
| 1.3.4 | | Dry the foil immediately using high-purity $N_2$ gas for 1–2 minutes. | | Drying: 1–2 minutes using a gentle, continuous flow | Handle carefully to avoid recontamination; ensure foil is completely dry before further processing. | |
| 1.4 | | Electropolishing (optional) | Concentrated phosphoric acid (85%); DI water; clean glass container; Electropolishing cell (glass beaker); copper cathode or Pt/Ti mesh; DC power supply; multimeter to monitor current; IPA; high-purity $N_2$ gas source with clean nozzle | | | Foil appears mirror-like and smooth, with significantly reduced surface roughness (target Ra ~180 nm or lower), verified by subsequent AFM or WLI. |
| 1.4.1 | | Prepare the electropolishing solution by diluting concentrated phosphoric acid (85%) with DI water in a 7:3 volume ratio in a clean glass container. | | Dilution: 7 parts acid to 3 parts water at room temperature (~25 °C) | Use appropriate PPE; ensure solution is homogeneous; prepare fresh solution if in doubt. | Electropolishing solution is clear and at the correct concentration as verified by volume measurements. |
| 1.4.2 | | Set up a two-electrode electropolishing cell with the Cu foil as the anode and a copper plate or Pt-coated Ti mesh as the cathode. | | Electrode separation: ~4 cm; use a stable setup ensuring uniform current distribution | Ensure electrodes are clean and securely positioned; avoid air bubbles trapped on the foil surface during immersion. | Electrode setup is stable and properly aligned as verified visually. |
| | | Maintain an electrode separation of approximately 4 cm. | | | | |
| 1.4.3 | | Apply a DC voltage of approximately 2.7 V across the cell and electropolish the foil for 450 seconds. Monitor the current to ensure stable operation. | | Voltage: ~2.7 V; Time: around 450 seconds (7.5 minutes); room temperature (~25 °C) | Over-voltage may cause burning or pitting; ensure voltage is stable; monitor for bubble evolution; replace solution if it becomes too saturated with Cu salts. We typically run this until the current is no longer decreasing significantly. | Bubbles will evolve and the foil will begin to look brighter in appearance as the roughness is decreased. |
| 1.4.4 | | Immediately remove the foil from the electropolishing cell and rinse thoroughly with DI water for 5 minutes to remove residual acid. | | Rinse time: 5 minutes; water at room temperature | Incomplete rinsing can leave residual acid; ensure thorough rinsing to prevent further chemical reactions on the foil surface. | Foil is free of acid residues; rinsing verified by a uniform appearance and lack of corrosion spots. |
| 1.4.5 | | Dip the foil briefly in IPA and immediately dry using a gentle, continuous flow of high-purity $N_2$ gas for 1–2 minutes. | | Dip in IPA for ~10 seconds; $N_2$ drying for 1–2 minutes | Handle carefully to avoid recontamination; ensure IPA is pure and free of particulates. | |
| 1.5 | | Controlled Oxidation of foil | A clean, uncontaminated hotplate with ±5 °C control; Cu foil from previous steps; | | | Uniform oxide thickness (consistent colour) across the entire foil; oxide layer thickness can be confirmed by cross sectional SEM if required. |
| 1.5.1 | | Preheat a hotplate with precise temperature control to 200 °C. Verify temperature stability with a thermocouple. | | Temperature: 200 °C; preheat until stable (typically 5 minutes) | Ensure accurate calibration of hotplate; avoid accidental contact with the hot surface. | Hotplate is uniformly heated to 200 °C as verified by thermocouple readings. |
| 1.5.2 | | Place the entire Cu foil onto the hotplate ensuring full contact (foil surface flat against the plate). Do not use any additional support to avoid uneven heating. | | Ensure complete and uniform contact with the hotplate; | Avoid sliding or moving the foil during oxidation; check that the foil is not warped or folded before heating. | Foil is uniformly in contact with the hotplate, confirmed visually. |
| 1.5.3 | | Heat the foil on the hotplate for 10 minutes to form a uniform oxide layer. Monitor the colour change on the foil surface. | | Heating duration: 10 minutes at 200 °C | Ensure even heating; do not overheat or underheat; slight colour change indicates oxide formation. | A uniform oxide layer (typically a consistent matte grey or light brown) is formed; validated by visual inspection. The foil should uniformly change colour at the same time at all places on the foil due to even oxidation. If some locations are not oxidizing, or some rings form, it is likely that there is some residual contamination present from either not cleaning thoroughly enough, or some intermediate contamination. If oxidation is not uniform, discard and start from 1.2 |
| 1.5.4 | | Remove the foil from the hotplate using clean tweezers and allow it to cool in a dust-free environment. | Clean tweezers; a clean cover to protect during cooling | Cool naturally to room temperature (approx. 25 °C); cooling time depends on ambient conditions, typically 10–15 minutes | Handle carefully as the foil may be hot; | |
| 2 | Reactor Pre-treatment | Prepare the reactor by flushing all gas lines to vacuum prior to opening valves, evacuating to a base pressure (~$4.2\times10^{-2}$ mbar). Then ramping the reactor temperature from room temperature to 1065 °C at ~50 °C/min with a mixture of 64:576 $H_2$:Ar. Allow a 30-minute warm-up at target temperature before cooling the reactor. | Aixtron Black Magic Pro 4" reactor; high-purity Ar gas; high-purity $H_2$ gas; mass flow controllers; vacuum pump | Ramp: ~50 °C/min; Target temperature: 1065 °C; Base pressure: ~$4.2\times10^{-2}$ mbar; operating pressure ~ 50 mbar; Warm-up: 30 minutes | verify stable sensor readings. | Stable, reproducible reactor conditions confirmed by consistent temperature and pressure readouts. |
| 3 | Calibration | Thermocouple calibration by melting Cu foil. Due to position of thermocouples which cannot be in contact with the sample without affecting growth, | Aixtron Black Magic Pro 4" reactor; high-purity Ar gas; high-purity $H_2$ gas; mass flow controllers; vacuum pump | | Temperature (or power of the heaters) should be increased step-wise and the system allowed to stabilise before increasing the temperature again. | The settings for the reactor should be correct and calibrated allowing for a consistent growth temperature, despite variations in thermocouple readings due to positioning. |

| | | temperature is calibrated by melting Cu foil, recording this temperature, and decreasing this by approximately 20C to reach appropriate growth temperature. This is done by following 2, but once temperature is stable, increasing until the Cu foil melts. | | | | |
|---|---|---|---|---|---|---|
| 4 | Sample Loading | Quickly load the pre-treated Cu foil onto the quartz or graphite holder on the heater with the growth side (front) facing upward. Align the foil flatly on the heater surface using clean tweezers. | tweezers; | Load immediately after pre-treatment or, if stored under vacuum, immediately after removal from vacuum; ensure complete flat contact with the heater; maintain ambient clean conditions | Do not fold or scratch the foil; | Even placement without folds or indentations; uniform thermal contact verified by visual inspection (it should lie flat on the surface of the heater) |
| 5 | Annealing and growth. | | High-purity Ar gas; reactor control system; mass flow controllers; High-purity $H_2$ gas; Methane ($CH_4$, we use diluted to 5% in Ar, with MFCs calibrated to Argon); | | The pressure should be held at 50 mbar during all annealing and growth processes: growth is sensitive to pressure, temperature, gas ratios and gas purity, varying any of these will change the growth parameters and resulting graphene film. | |
| 5.1 | | Ar anneal stage: Ramp the reactor under high-purity Ar to 1065 °C at ~50 °C/min, then hold at 1065 °C for 30 minutes at 50 mbar and 640 sccm Ar to promote grain growth and reduce impurities. | | Ramp rate: ~50 °C/min; Hold: 30 minutes at 1065 °C, 50 mbar, 640 sccm Ar | Ensure uniform temperature distribution – any variation from the usual profile may indicate that something has shifted in the system and the system must be reconfigured and calibrated (beyond the scope of this STEP) | Formation of smoother, larger Cu grains with visible thermal grooving; confirmed by optical microscopy or EBSD. |
| 5.2 | | $H_2$ anneal stage: Introduce high-purity $H_2$ gas along with Ar in a flow ratio of ~64:576 sccm at 50 mbar. Continue annealing at 1065 °C for a total of 60 minutes to further reduce surface oxides. | | $H_2$:Ar ratio: 64:576 sccm; Total annealing time: 60 minutes at 1065 °C; Pressure: 50 mbar | Hydrogen is flammable; ensure proper ventilation and leak detection; maintain stable gas flows and reactor pressure; | Reduction of residual oxides confirmed by lower oxide signals (via ToF-SIMS or XPS) and increased nucleation uniformity in subsequent graphene growth. |
| 5.3 | | Introduce the carbon precursor by flowing diluted $CH_4$ along with $H_2$ and Ar at 1065 °C. Maintain precise gas flow ratios to achieve low supersaturation and allow growth of large graphene domains. | | Gas flow ratio: $CH_4$:$H_2$:Ar = 0.32:64:576 sccm; Temperature: 1065 °C; Pressure: 50 mbar; Growth time: 60 minutes | $CH_4$ is flammable; perform leak tests; ensure valves are shut when not in use to prevent MFC leakage; monitor for trace oxygen contamination. | Continuous monolayer graphene film; reproducible nucleation density and domain sizes; validated by a short 1-minute hotplate test at 200 °C on corner of Cu foil, showing minimal Cu oxidation. |
| 5.4 | | Cool the reactor and sample under continuous high-purity Ar flow. Allow the temperature to drop gradually to below 200 °C within ~1 hour. Vent the reactor only after the foil is near room temperature (<50 °C). | | Cooling time: ~1 hour to below 200 °C; Ar flow maintained at ~500 sccm; vent only after <50 °C | Maintain inert atmosphere to prevent oxidation; | Foil reaches room temperature with absence of oxidation confirmed by optical microscopy. If there is oxygen present in the Ar, it may be seen that graphene has many small hole, or if there is significant O2 contamination, crystalline cuprous oxide might form in any region where there is no graphene. If this is the case, repeat this guide after oxygen has been sufficiently removed from the gas lines. |
| 6 | Extraction/ Removal | Remove the Cu foil with grown graphene from the reactor using clean tweezers. Handle the foil only at the corners and transfer it to a clean container immediately. | Tweezers; appropriate storage container (Polypropylene containers are found to be sufficient) | Extraction only after complete cool-down; handle gently; avoid mechanical stress by grabbing only at the corners | Caution due to residual heat; minimize mechanical stress to prevent damage to the graphene film. | Monolayer (<3% multilayers) graphene on Cu foil with full coverage. Further characterisation can be done, e.g. Raman and SEM, however, these characterisation protocols are beyond the scope of this STEP. |
| | | | | | | |

## 5. Example of STEP for transfer of CVD graphene from Cu foil

The example presented here shows a procedure for transferring CVD graphene from copper foil onto a 90 nm SiO2/Si substrate, using chemical etching wet transfer. The procedure uses Ammonium Persulfate as the etchant; however the process is also applicable for other etchants used in literature (e.g. – $FeCl_3$, $HCl/H_2O_2$, $HNO_3$). The procedures are also generally relevant to the etching transfer of other 2D material transfer systems (e.g. – hBN/Cu, TMD/Au) as well as electrochemical transfer methods.

This example took between 4-6 hours to complete.

*Supplementary Table 4. Example of STEP protocol for transfer of CVD graphene from Cu foil.*

| # | Main task | Sub task | ME – Materials and Equipment | PR – Parameters and Ranges | IWTD – Issues, Warnings, Troubleshooting and Difficulties | VEO – Validation and Expected Outcomes |
|---|---|---|---|---|---|---|
| **Description of checkpoints** | | | Specify all materials and equipment used, including alternatives if primary options are not available. Include details about manufacturers, models, and modifications. | List and record all controlled and uncontrolled parameters. Provide a range of values tested for each critical parameter. | Identify potential safety hazards, operational issues, and provide troubleshooting tips. Point out if and how a step is difficult, and how the experimenter can reduce the difficulty to increase the chance of success. | Describe the expected results or outputs clearly, including any specific observations or measurements that indicate that the process/characterisation step was successful. |
| 1.1 | Polymer Solution Preparation (10% w/w PMMA 966k in Anisole) | Prepare a 250 ml clean glass bottle with a cleaned PTFE magnetic stirring bar. If not brand new, the PTFE stirrer is wiped with IPA-soaked lint-free tissue to remove particulate dust and potential remaining residues of previous solutions.<br><br>Ensure the bottle has been thoroughly cleaned prior to use and is completely dry. | Materials: VWR PTFE magnetic stirring bars; Duran borosilicate glass bottle, 250 ml | Controlled parameters: Visible dust/particles, liquid droplets or wetness in the glass bottle, residues or discoloration on the stirring bar. | PTFE stirrer: propensity for collecting dust due to static buildup, if these end up in the solution they will create inconsistencies in the coated film in subsequent steps; if the stirrer is for general use in the lab, it may have collected residues from previous uses. It is also important to use a stirrer of the appropriate size for the solution and bottle.<br><br>Glass bottle: If the glass bottles have been used to store solutions prior to use, even if it is the same solution, there is a risk of contaminating the prepared solution. It is important to thoroughly clean and dry the glass bottle and rinsing it with the solvent to be used can provide further reassurance that the bottle is thoroughly clean. | Clear glass bottle without any solvent or particulate residues; PTFE stirring bar free of particulate or polymer residue contaminants. |
| 1.2 | | Pour 225g Anisole into the glass bottle, either by pre-weighing the solvent in another clean glass beaker or pouring it into the glass bottle as it sits on a weighing scale. | Equipment: any weighing scale (+/0.1g or better) will do.<br><br>Materials:<br>Anisole, anhydrous 99.7%, Sigma-Aldrich #296295-1L | Controlled parameters:<br>- Anisole weight: +/- 1g<br>- anisole purity<br>- solvent used (Anisole)<br><br>Uncontrolled parameters:<br>- any invisible residues in glass beaker (if used) | Risk of spilling anisole during pour, risk of contamination from glass beaker (if not properly cleaned). Conduct in a fumehood or properly ventilated area, or as required by safety procedures. | Visually clear liquid in bottles, free of particulate contamination. |
| 1.3 | | Gently add 25 g 966k PMMA into the glass bottle as it sits on the weighing scale. Add small amounts of powder at a time to prevent agglomeration at the bottom. | Equipment: any weighing scale (+/0.1g or better) will do.<br><br>Materials:<br>PMMA, Mw 966,000, Sigma-Aldrich #182265-500G | Controlled parameters:<br>PMMA weight: +/-1g<br>PMMA molecular weight<br>PMMA quantity (10% w/w) | Risk of spilling PMMA powder on weighing scale or around the work area, giving erroneous readings on the weight of solution; risk of PMMA powder agglomerating in the solvent as it sinks to the bottom.<br><br>These can be mitigated by adding small amounts of powder at a time, and stirring the solution periodically to disperse the powder. | Limited agglomeration or clumping of PMMA powder at the bottom of the bottle. |
| | | | | | Conduct in a fumehood or properly ventilated area, or as required by safety procedures (Handling Anisole). | |
| 1.4 | | Close the lid on the glass bottle and place in on a stirring plate. Stir solution for 24 hours at 500-1000 rpm. Continue stirring for longer if any visible particles of powder remain. | Equipment: any stirring plate will do. | Stirring speed: 500 rpm – 1000 rpm based on size of the bottle and stirring bar.<br><br>Duration of stirring: 24 hours. Longer durations ensure that microscopic powders are also thoroughly dissolved in solution. The duration can be made longer, e.g. – 48 hours. | Ensure the powder has thoroughly dissolved in solution. Closing the lid on the bottle ensures the solvent does not evaporate during prolonged stirring.<br><br>We do not heat the solution during stirring to avoid any evaporation of the solvent or pressure buildup in the bottle. Anisole is also flammable.<br><br>Once a 250 ml solution is made, it can be used for multiple transfers; however, depending on the storage conditions (humidity, sunlight, temperature) the polymer can degrade and solvent evaporate over time, so the properties of the solution can vary the older it is. Ideally, the mixture should be used before 6 months and periodically may need to be stirred to ensure contents do not sediment. Use a fresh disposable pipette everytime to draw solution from bottle; this ensures the solution remains contamination-free.<br><br>Conduct in a fumehood or properly ventilated area, or as required by safety procedures (Handling Anisole). | Visually clear solution without visible particulate matter or bubbles. |
| 2 | Transfer piece preparation | Section graphene-coated copper foil to desired dimensions – 5 x 5cm2 pieces using IPA-cleaned sharp shears or scissors. | Precision shears, cutting guide/template, or scissors. | We prepare a larger piece for polymer coating, and after the coating process, we section the foils to sizes we wish to transfer to the sample substrates. | Take great care in avoiding creases or wrinkles in the copper foil; these areas will not yield a successful graphene transfer. Use the cutting tool with which the user is most able to avoid mechanical deformation.<br><br>Avoid touching the graphene-coated surface with metallic, ceramic or other sharp objects as these will damage the graphene.<br><br>Handle the foils from the edges wherever possible with tweezers and wear nitrile gloves when handling the foils; it is acceptable to rest the backside of the foil on the palm of one's hand (wearing gloves) if this facilitates easier handling.<br><br>It may be useful to cut a larger piece and subsequently to cut smaller pieces once the polymer coating is applied. This would provide a more even coat over the foil surface and reduce the number of cuts before the graphene is protected. | Copper foil is cut to dimensions and is devoid of any mechanical deformation in the target transfer area. |
| 3.1 | Coating PMMA carrier film on transfer piece | Secure the transfer piece on a solid surface such as a silicon wafer. Ideally, the wafer is coated with PDMS, onto which the foil can be attached.<br><br>Alternatively, tape all four sides of the foil | Materials: Solid, flat substrate (e.g. – silicon or quartz wafer), Kapton tape (if needed)<br><br>We use a silicon wafer with a cast PDMS film (Sylgard 184) on top as a support (2-inch wafer or 4-inch wafer variants depending on the size of the sample). | Flatness of the copper foil on substrate, how well the edges are sealed/taped, which type of tape is used (Kapton tape). | There is a risk that the foil is mechanically deformed during mounting, or that the foil is not seated flat against the substrate. Ensure that the foil is completely adhered to the PDMS, or alternatively that the edges are fully taped. The polymer solution will coat the backside of the foil if there | Transfer piece is mounted flat against the substrate, and there are no visible gaps between the foil and substrate on the edges. Foil is free of creases or wrinkles. |
| | | onto the substrate using Kapton tape or similar. Do not use scotch tape as it can interact with the anisole solvent. Ensure that the foil is seated flat against the substrate, and that the foil does not mechanically deform during mounting. | | | are any gaps in between the foil and the substrate, complicating subsequent processing. | |
| 3.2 | | Mount the substrate centered on the spincoater chuck. Engage vacuum and run a test recipe to gauge centering on chuck and adhesion of the foil onto the substrate. | Equipment: any spincoater will do.<br><br>Materials: we use a custom-made centering tool to help position the wafer on the vacuum chuck. | Rotation speed: 1500 rpm<br>Acceleration: 500 ms<br>Duration: 30 seconds | Centering of the substrate may be off, which may lead to inhomogeneous coating or the substrate detaching during spinning. Adjust centering as needed until the substrate is visibly centered when spinning.<br><br>If the foil has been mounted onto PDMS, and it is peeling off during spinning, then either the PDMS needs to be cleaned, or that the foil is too large. Clean PDMS and retry or tape the edges of the foil. | Substrate is visibly centered on the chuck while spinning, and the foil does not delaminate during spinning. The substrate stays on the vacuum chuck during spinning. |
| 3.3 | | Coat the copper surface with polymer solution using a polypropylene disposable pipette, such that the liquid coats the entire foil surface.<br><br>Spincoat at 1500 rpm for 60 seconds, using an acceleration time of 500 ms. | Equipment: any spincoater will do.<br><br>Materials: polymer solution previously made; propylene disposable pipette | Rotation speed: 1500 rpm<br>Acceleration: 500 ms<br>Duration: 60 seconds<br><br>Uncontrolled variable: How much polymer solution is dispensed on the surface. We put just enough liquid such that the entire surface is wetted, but not overflowing, with liquid. | Conduct in a fumehood or properly ventilated area, or as required by safety procedures (Handling Anisole).<br><br>Put just enough liquid to wet the entire surface, but not flood it. Use the disposable pipette to spread the liquid around on the surface but ensure the pipette does not touch the copper surface when doing so.<br><br>By controlling the rotation speed, a thinner PMMA film can be used. We find that thinner films have a higher chance of disintegrating during etching or handling, so we opt for the thicker films to improve yield. | A uniform coating is visible on the copper surface, and there are no liquid droplets on the foil (apart from edges). |
| 3.4 | | Dismount the foil from the substrate gently, taking care to avoid creating the foil. Place the foil on a hot plate and perform a baking step.<br><br>We place the foil on a hot plate at room temperature and then set the temperature to 165˚C. Once the temperature reaches 165˚C, we bake the sample for 5 minutes. We remove the foil from the hotplate afterwards and let it cool to room temperature.<br><br>We then make a chiral mark on top of one of the edges of the PMMA-coated side using a permanent marker, so we can keep track of which side of the transparent polymer the graphene is on (we use the word 'TOP' or the number '4' – if we are facing the side where we can read 'TOP' or '4', then the graphene is on the opposite side of the polymer. | Equipment: Any hotplate with precise temperature control (plus minus 5 degrees) will do. | Temperature: 165˚C<br>Bake time: 5 minutes<br><br>Uncontrolled variable: we do not control the ramp time to 165˚C. | Conduct in a fumehood or properly ventilated area, or as required by safety procedures (Handling Anisole).<br><br>If there were exposed edges during spincoating, there may be polymer residues on the back side which will need to be removed prior to subsequent steps. Rinse the backside with solvent from a dispensing bottle or disposable pipette, ensuring the foil is not mechanically deformed in the process, or that the front side is not accidently washed with the solvent. | Upon visual inspection, there appears to be a thin homogeneous coating on the copper foil, without any particles or holes in the coating. |
| 3.5 | | Remove graphene on the backside of the copper foil: place the | Equipment: Plasma ashing system PE-50 from Plasma Etch. | Plasma Power: 50W<br>Plasma Duration: 30-60s | If there is any polymer on the backside of the foil, it will not come off in this step. | |
| | | foil face down in the plasma asher. Plasma treat the backside to remove the graphene.<br><br>We use 60s at low power $O_2$ plasma (50 W, at medium-high pressure, 50 mTorr), O2 only flow (flow adjusted to maintain 50 mTorr). | | O2 flow rate: adjusted to maintain constant pressure.<br><br>Plasma parameters can vary considerably between system and between runs. We optimize the process parameters to ensure there is no graphene left on the backside, but do not go to the extent that the copper becomes heavily oxidized (beyond a red color). | If there is any graphene left on the backside, it will slow down the copper etching process and will adhere to the transferred side once the copper is fully etched.<br><br>The foil can flip inside the plasma chamber while flushing/purging. Secure the edges of the foil if needed using a heavy solid object, such as a glass slide. | |
| 4 | Etchant solution preparation | We use a solution of 1M ammonium persulfate, and a quantity of roughly 50 ml of solution for a 2x2 cm piece. We make a stock solution of 1L and use it as needed for transfers, using the solution within 6 months of preparation.<br><br>The solution is prepared by pouring 500 ml Milli-Q® water into a 1L glass bottle (cleaned and dried, and containing a cleaned PTFE magnetic stirring bar), add 228g of ammonium persulfate, and subsequently add enough water to reach the 1L mark on the bottle. | Materials:<br>Ammonium persulfate ≥98.0% , Sigma-Aldrich #248614-500G;<br>Milli-Q® water;<br>Duran borosilicate bottle, 1L | Ammonium persulfate molarity in solution: 1M<br><br>Amount of solution used per transfer: approximately 50 ml for a 2x2 cm piece<br><br>We make the etching solution in batches and use it as needed for transfers. Solutions typically last about 6 months at a time, but the age of the solution is not strictly controlled. | Risk of spilling ammonium persulfate powder on weighing scale or around the work area, giving erroneous readings on the weight of solution; risk of powder agglomerating in the solvent as it sinks to the bottom.<br><br>These can be mitigated by adding small amounts of powder at a time and stirring the solution periodically to disperse the powder.<br><br>We do not heat the solution during mixing. | Clear solution without sediments at the bottom of the bottle, or dust/particles/bubbles circulating in solution, at the bottom or on the surface of the solution. |
| 5.1 | Graphene etching transfer | We pour the requisite amount of solution into an appropriately sized glass beaker, and float the transfer piece on top of the solution. Ensure that the PMMA-coated side is not submerged in the solution.<br><br>We cover the solution with Parafilm®M and leave it in the fumehood unperturbed, until the copper completely dissolves. | Materials: Etchant solution previously made; transfer piece; Parafilm®M; glass beaker. | We also do not strictly control the length of etching, as it can vary between samples. In general, the sample is left for a couple of days until the copper fully disappears. | Sometimes the copper does not fully etch in solution even after a few days. In this case, change out the solution with fresh etchant.<br><br>Avoid flipping or submerging the foil when etching, as it can be tricky to try to flip the foil back once the copper is dissolved. It is also important to avoid getting droplets on the top surface of the PMMA film as these can be tricky to remove in subsequent steps or may lead to etchant contamination in the final sample.<br><br>Thinner PMMA films can sometimes disintegrate in solution or during subsequent handling, which is why opt for the thicker films used in our protocol. | Copper foil is visibly fully dissolved, and the solution has turned a pale blue color. There are no cupric residues attached to the floating PMMA film. The PMMA film is intact and has not disintegrated. |
| 5.2 | | Once the copper foil has completely dissolved, we prepare to remove the transfer piece from the etching solution and to rinse it.<br><br>We prepare 3 Milli-Q® water baths of roughly 200 ml each for a 2x2cm piece. After removing the Parafilm®M cover, we use a microscope glass slide to gently scoop out the transfer piece.<br><br>If there are any solution residues on the top of the PMMA film, we then gently rinse the top surface | Materials: glass slide(s); Milli-Q® water; 3 glass beakers; lint-free tissue. | The hold-time in each bath is not strictly controlled, but approximately around 2-3 minutes per bath. Longer bath times are generally preferable (depending upon one's patience) as they give more time for etchant contaminants to be removed in each bath. However, rinsing in 3 baths also largely removes these contaminants.<br><br>The amount of water in each bath is also not strictly controlled. More water is preferrable, however, | If there is any residue on the top side of the PMMA film, it needs to be rinsed and removed before placing it in the Milli-Q® water baths, because this residue will carry over to the final transfer.<br><br>We do not heat our water baths, but there are some groups that do, and this can help reduce contamination if there is an issue in the final device.<br><br>Rinse the glass slides between bath changes or use a new slide to transfer the film to the next bath. The glass slides will carry over residues as well if not rinsed. | A transparent, clear film is floating atop the 3rd bath, and there is no visible blue tint in the 2nd and 3rd baths. There is also no visible copper residues (orange, red or blue) on the film, nor are there any droplets atop the film. |
| | | while it is on the glass slide with Milli-Q® water, taking care not to accidently flip the film. We remove any excess water droplets on the topside of the film with lint-free tissue.<br><br>We then gently float the transfer piece atop the 1st Milli-Q® water bath for 2-3 minutes and then scoop it out again with the same glass slide. We repeat the process with the 2nd and 3rd baths and keep the film afloat on the 3rd bath prior to the next step. | | | There is a risk that the PMMA film disintegrates, or that it flips over. While nothing can be done if the PMMA disintegrates, it may still be possible to flip the film back to the correct side with trial and error. However, there is a chance that the film disintegrates in the process. Using a thicker PMMA film reduces the chance for flipping and disintegration in this crucial stage. | |
| 6.1 | Sample preparation | Section a suitable size of $SiO_2$ piece using a diamond scriber. We place the $SiO_2$ wafer face down on lint-free tissue atop an anodized aluminum slab. We scribe the backside of the wafer and cleave it using the edge of the slab. The wafer is moved to a clean part of the tissue each time it is cleaved to ensure silicate particles do not damage the surface. | Materials: diamond scriber, a hard surface (aluminum slab or similar), target piece (90 nm $SiO_2$/Si wafer in this case). | Dimensions of target piece: 3x3cm in this case (for a 2x2cm film).<br><br>We avoid placing the wafer or cut pieces face down in areas on the lint-free tissue where there are silicate residues visible. These will scratch the surface. | Risk of generating silicate particles during cleaving; it is important to protect one's eyes with safety goggles.<br><br>Risk of damaging gate oxide or scratching the surface if cleaving and scribing is done on the same spot on lint-free tissue. | Target pieces are cut to dimensions slightly larger than the transfer piece (e.g. – 3x3 cm target piece for a 2x2cm transfer piece). |
| 6.2 | | Rinse the sample surface with acetone and IPA from a dispensing bottle and immediately dry the surface with a $N_2$ gun.<br><br>As an option, the sample can be additionally surface treated to improve adhesion. We plasma treat the surface of $SiO_2$ chips in an $O_2$ plasma at 300W for 2 minutes, 50 mTorr pressure. | Equipment: Plasma ashing system PE-50 from Plasma Etch.<br><br>Materials: Acetone, IPA (dispensing bottles), $N_2$ gun, target piece. | Controlled parameters:<br>Plasma power: 300W<br>Plasma time: 2 minutes<br>$O_2$ flow rate: adjusted to maintain 50 mTorr pressure.<br><br>Uncontrolled parameters: length of rinsing, purity of Acetone and IPA.<br><br>Parameters may vary considerably depending on the system as well as the substrate. Optimisation of parameters should be done on local system and substrate, to balance stickiness with damage to substrate. | The purpose of this step is to remove any dust and particulate residues from the cleaving process. It is also to improve graphene adhesion to the surface – which may or may not be needed depending upon environmental or substrate surface conditions. | No visible particles or residues on the substrate surface. |
| 7.1 | Graphene transfer | Once the sample has been cleaned, we immediately proceed to transferring the graphene.<br><br>We scoop out the transfer piece from water gently on top of the sample, ensuring that the film does not have any wrinkles when it is lying on the sample. If there are wrinkles present, we redo the fishing process. | Materials: target piece on which to transfer graphene. | If the substrate has been plasma-cleaned, we aim to perform transfer as fast as possible (< 2minutes between when the substrate is out of the plasma tool and into the water). | Ensuring that the PMMA film does not wrinkle, flip or disintegrate, or that there are no bubbles between film and substrate, will largely dictate how well the graphene is transferred.<br><br>Repeat the scooping process until a smooth, wrinkle-free film is achieved on the substrate. | Smooth, wrinkle-free film is transferred onto the target substrate. |
| 7.2 | | Once the film is on top of the sample – without wrinkles – we let the sample air dry until there is no visible liquid between transfer piece and sample surface – this typically takes 10-20 minutes.<br><br>We place the sample on a hot plate at room temperature and then set the temperature to | Equipment: Any hotplate with precise temperature control (plus minus 5 degrees) will do. | Controlled Parameters:<br>Drying temperature: 50˚C<br><br>Uncontrolled parameters:<br>Time to air-dry<br>Time to dry at 50˚C<br><br>We do not control these parameters as they vary significantly between samples, so we instead visually inspect the samples. | How the sample is dried is critical to avoiding wrinkles, creases, or bubbles in the PMMA film, and ensuring proper coverage following transfer. This step will heavily influence transfer success.<br><br>The purpose of the 50˚C drying step is to remove any additional water not visible to the naked eye, and to | A smooth film is coated on top of the substrate, and there are no visual defects such as wrinkles, creases, trapped bubbles, or delamination of the PMMA film from the surface. |
| | | 50˚C to allow for a gently drying out of the film before baking the sample. | | | promote movement of any trapped water/air bubbles. This step ensures a more seamless adhesion of the PMMA film to the substrate.<br><br>Leave the sample on a lint-free tissue for initial drying. If it is placed | |
| 7.3 | | We ramp the temperature to 165˚C and, once it is at this temperature, bake the sample for 5 minutes. We then remove the sample from the hot plate and let it cool down to room temperature. | Equipment: Any hotplate with precise temperature control (plus minus 5 degrees) will do. | Controlled Parameters:<br>Bake temperature: 165˚C<br>Bake time: 5 minutes<br><br>Uncontrolled parameters: Cooling speed to room temperature. Given how small the sample is, it cools almost immediately when placed on a lab bench. | Risk of the PMMA film delaminating from the surface, which is a sign the adhesion to the surface is poor. Plasma treating the surface may be required. | A smooth film is coated on top of the substrate, and there are no visual defects such as wrinkles, creases, trapped bubbles, or delamination of the PMMA film from the surface. |
| 8.1 | Polymer Removal | We use 100 ml acetone to dissolve PMMA. The sample is placed in a beaker containing 100 ml acetone and is left there for 30 minutes – 1 hour. | Materials:<br>Acetone 99,8, HiPerSolv CHROMANORM® HPLC, VWR Avantor # 20067.320; Glass beaker | We do not strictly control the amount of acetone used, or the removal time. Typically, an excess of acetone is used compared to the sample size. PMMA typically dissolves within 5 minutes, so the longer duration is simply to ensure all the PMMA has dissolved. | Heating the acetone solution may be helpful in reducing the amount of residues left from the polymer, but we do not do this for our standard process. If we observe contamination to be an issue, then we heat the solution in subsequent transfers.<br><br>One may also choose to leave the sample in acetone overnight, and to heat the acetone in the meanwhile as well. In this case, it is important to ensure there is enough solvent in the beaker, and to cover the beaker with 2x aluminum foil to reduce evaporative loss.<br><br>On some substrates, the graphene will delaminate from the surface during this step. In this case, mount the substrate on a glass cover, place it on an acetone solution, and heat the solution. Acetone vapors will dissolve the PMMA without causing the graphene to delaminate.<br><br>When working with solvents, operate in a fumehood or ventilated area, or as required by safety procedures. | PMMA film is visibly removed from the substrate surface. |
| 8.2 | | Once the PMMA film has fully dissolved in acetone, we take the sample out of the solution, rinse it with IPA from a dispensing bottle, and then immediately dry the sample using an N2 gun, before the IPA has a chance to dry out. We dry both sides of the sample and continue drying until the sample is free of any liquid on either side. | Materials: IPA (dispensing bottle), N2 gun, lint-free tissue (to rest the sample on after/during drying). | We do not strictly control the rinsing time with IPA, and the N2 drying time is adjusted based on whether there are any solvent residues left on the surface. | It is important the dry the substrate surface with the N2 gun, and to not allow the solvent to evaporate on its own – this will leave coffee-ring residues on the substrate which are difficult to remove.<br><br>Ensure that both sides of the substrate are fully dried before placing the substrate on a flat surface – any residual solvent will cause the substrate to stick to the surface.<br><br>When working with solvents, operate in a fumehood or ventilated area, or as required by safety procedures. | There are no 'coffee ring' solvent residues on the substrate, and optical microscopy images of the sample surface confirm significant residues of PMMA are not present. |

## 6. Supplementary Information References